%% file: main-arxiv.tex
\documentclass{article}

\usepackage{arxiv}

\usepackage[utf8]{inputenc} 
\usepackage[T1]{fontenc}    
\usepackage{hyperref}       
\usepackage{url}            
\usepackage{booktabs}       
\usepackage{amsfonts}       
\usepackage{nicefrac}       
\usepackage{microtype}      
\usepackage{lipsum}		
\usepackage{graphicx}
\usepackage{natbib}
\usepackage{doi}

\usepackage{siunitx}
\usepackage{booktabs,caption}
\usepackage{hyperref}
\usepackage{indentfirst}
\usepackage{subcaption}
\usepackage{float}
\usepackage{bm}
\usepackage{color, colortbl}
\usepackage{array}
\usepackage{placeins}
\usepackage{amsmath}
\usepackage{xcolor}
\usepackage[normalem]{ulem}
\usepackage{tabularx}
\usepackage{tikz}
\usetikzlibrary{positioning,fit,arrows.meta,calc}

\newif\ifrevmarks
\revmarkstrue
\ifrevmarks
  
  \newcommand{\del}[1]{\textcolor{red}{\sout{#1}}}
  \newenvironment{revblock}{\color{red}\captionsetup{font+={color=red}}\captionsetup[sub]{font+={color=red}}}{}
  \newcommand{\delfloat}[1]{#1}
\else
  
  \newcommand{\del}[1]{}
  
  \newcommand{\delfloat}[1]{}
\fi

\newcommand{\btheta}{\bm{\theta}}

\newcommand{\bbeta}{\bm{\beta}}
\newcommand{\blambda}{\bm{\lambda}}
\newcommand{\bX}{\bm{X}}
\newcommand{\by}{\bm{y}}

\title{\normalfont\bfseries\Large From priors to performance: enhancing statistical efficiency with Bayesian dynamic borrowing}

\author{
  \textbf{Xinxin Chen}\textsuperscript{1*} \and
  \textbf{Ezequiel Braga Santos}\textsuperscript{2*} \and
  \textbf{Joseph G.~Ibrahim}\textsuperscript{1} \and
  \textbf{Luiz Max Carvalho}\textsuperscript{1,2*\dag} \and
  \textbf{Ethan M.~Alt}\textsuperscript{1*}
}

\hypersetup{
pdftitle={From priors to performance: enhancing statistical efficiency with Bayesian dynamic borrowing},
pdfsubject={Bayesian dynamic borrowing with historical data},
pdfauthor={Xinxin Chen, Ezequiel Braga Santos, Joseph G. Ibrahim, Luiz Max Carvalho, Ethan M. Alt},
pdfkeywords={historical data, power prior, dynamic borrowing, model selection, average treatment effect},
}

\begin{document}
\maketitle

\let\thefootnote\relax\footnotetext{\textsuperscript{1}Department of Biostatistics, University of North Carolina at Chapel Hill, Chapel Hill, North Carolina,  27599-7420, United States of America}
\let\thefootnote\relax\footnotetext{\textsuperscript{2}School of Applied Mathematics, Getulio Vargas Foundation, Rio de Janeiro, Rio de Janeiro, 22250-900, Brazil}
\let\thefootnote\relax\footnotetext{\textsuperscript{*}Equal contribution}
\let\thefootnote\relax\footnotetext{\textsuperscript{\dag}Corresponding author: lmax.fgv@gmail.com}

\begin{abstract}
	One of the main advantages of the Bayesian approach to statistical inference is the flexibility in incorporating information from various sources, from expert opinion to historical data.
    Whereas the literature on Bayesian dynamic borrowing is rich, there is a relative scarcity of worked examples that apply several dynamic borrowing priors to the same individual-level data, from model specification to sensitivity analysis.
    We review the most common dynamic borrowing approaches for external data, including how to incorporate multiple historical data sets.
    We then show how these techniques can be used in practice, using publicly available software, to perform important statistical tasks such as model selection, average treatment effect estimation and prior sensitivity analysis.
    The aim is to make analysts more confident in their use of Bayesian methods to incorporate external data.
    All code is made publicly available in a companion GitHub repository (\url{https://github.com/EzequielEBS/hdbayes-tutorial}).
\end{abstract}

\keywords{historical data \and power prior \and dynamic borrowing \and model selection \and average treatment effect}

\input{001-intro}
\input{002-models_hist_data}
\input{003-applications}
\input{003_4-PSM}
\input{004-discussion}

\section*{Conflicts of Interest}
The authors declare no potential conflicts of interest.

\section*{Data Availability Statement}
All data and code used in the analyses presented in this paper are available at \url{https://github.com/EzequielEBS/hdbayes-tutorial}.
Claude (Anthropic; most recently Claude Opus 5.5) was used for language editing and code formatting, to check references and typesetting, to propose wording for some revised passages, and to draft the TikZ code for Figures~\ref{fig:borrowing_schematic} and~\ref{fig:workflow}.
The authors reviewed and edited all output and take full responsibility for the content of the manuscript.

\newpage
\bibliographystyle{unsrtnat}
\bibliography{tutorial_historical}

\newpage

\appendix

\setcounter{table}{0}
\setcounter{figure}{0}
\renewcommand{\thetable}{S\arabic{table}}
\renewcommand{\thefigure}{S\arabic{figure}}

\input{supp}

\end{document}

%% file: 001-intro.tex
\section{Introduction}
\label{sec:intro}

While the incorporation of external data in statistical analyses dates back many decades~\citep{Pocock1976}, rising costs and logistical challenges of collecting data prospectively have spurred interest in methods that leverage external data to augment contemporary studies~\citep{Lesaffre2024,Han2025}.
The Bayesian framework offers a natural solution by explicitly incorporating external information through informative priors, which can reduce sample sizes, increase statistical power, and, in some cases, render trials feasible when recruitment is difficult, e.g., in pediatric trials~\citep{Lesaffre2024}.
These approaches are particularly valuable in pharmaceutical research, where historical control arms, real-world data, or bridging studies are often available but may differ in patient populations or trial conditions~\citep{Viele2014}.
Recent guidance from the U.S. Food and Drug Administration (FDA) has also recommended the use of Bayesian methods in the analysis of clinical trials~\citep{FDA2026bayesian,Lee2026}.
The FDA instructs that the analysis of clinical trial data may include external data from previous trials and also ``incorporate features such as down-weighting of earlier data relative to contemporaneous data or initial skepticism regarding the likelihood of large treatment effects''~\citep{FDA2022}.

On the other hand, indiscriminate use of historical data can lead to bias and type I error inflation if there is conflict between historical and current data, which motivates the development of methods that dynamically modulate the influence of past evidence based on its compatibility with new data~\citep{Ibrahim2015,Jiang2023,altLEAP2024,Wang2025}.
In particular, \textbf{dynamic borrowing priors} (DBPs) -- such as the normalized power prior~\citep{Duan2006,Carvalho2021}, the robust meta-analytic-predictive (MAP) prior~\citep{Schmidli2014}, the elastic prior~\citep{Jiang2023}, the self-adapting mixture prior (SAM)~\citep{Yang2023}, multisource exchangeability models (MEMs)~\citep{Kaizer2018}, and the latent exchangeability prior (LEAP, \cite{altLEAP2024}) -- address this challenge by adaptively weighting historical information.
See recent reviews by~\cite{Lesaffre2024} and~\cite{Chen2024}.

Unlike static borrowing approaches (e.g., fixed-effect meta-analysis), dynamic methods incorporate and quantify the degree of similarity between historical and current data, down-weighting or discarding incongruent sources.
For instance, the normalized power prior~\citep{Duan2006,Ibrahim2015}  discounts the likelihood for the historical data \textit{via} a discounting parameter $a_0 \in [0, 1]$, which is estimated in a fully Bayesian way by assigning it a prior (i.e., a \emph{hyperprior}) and conducting posterior inference~\citep{Ibrahim2015,Carvalho2021}.
The MAP prior, on the other hand, assumes exchangeability across studies and estimates between-trial heterogeneity~\citep{Neuenschwander2010}.
It is also possible to use the meta-analytic framework with random effects to leverage the shrinkage properties of hierarchical models to obtain more stable estimates while taking heterogeneity into account~\citep{Rover2020} or to use variance-inflating \textit{elastic functions} to correct for discrepancies between historical and current data~\citep{Jiang2023,Huang2025}.

Despite their promise, dynamic borrowing methods pose unique challenges.
For example, regulatory agencies emphasize controlling type I error inflation when historical data are incorporated, and the choice between methods depends on the number of historical studies, model complexity, and computational trade-offs~\citep{Schmidli2014}.
Recent advancements, such as robustified priors and hybrid propensity-score approaches, further refine these methods in an attempt to mitigate prior-data conflict~\citep{Banbeta2019}.
In light of these advances, we feel it is important to provide practical guidance for how to implement various DBPs within a modern Bayesian data analysis, including issues of prior calibration, prior sensitivity, model selection and prediction.
In this sense, our approach is complementary to that of recent work such as \citet{Lesaffre2024}, which gives a broad overview of the literature and its connections, as well as the implications of using DBPs from a regulatory perspective.
Software-oriented guidance is available for specific priors, for example for the MAP prior in the R package \textbf{RBesT}~\citep{Weber2021}, and \citet{Lee2024} gives a tutorial review of Bayesian methods for confirmatory trials in the regulatory setting.

This paper showcases the philosophy, applications, and practical considerations of dynamic borrowing through the lens of a modern Bayesian analysis, with a focus on Bayesian methods for the \textit{analysis} of historical data settings.
Readers interested in \textit{design} are directed to \cite{Ghadessi2020} and \cite{Wang2025b}.
We show how to use DBPs to perform model selection and estimate causal effects for generalized linear models and survival analysis using real data. 
Many of the computations in the present paper rely on the \textbf{hdbayes} R package, described in~\citet{Alt2025}, and we revisit some of the priors and examples in that paper.
We substantially extend~\citet{Alt2025} in several directions, giving practical guidance on Bayesian $g$-computation, prior sensitivity analysis, prior predictive checks, leave-one-out cross-validation using \textbf{loo}~\citep{loo_package}, and propensity score matching combined with dynamic borrowing (using \textbf{psrwe}~\citep{psrwe_package}, \textbf{dbarts}~\citep{dbarts_package} and \textbf{MatchIt}~\citep{hoMatchItNonparametricPreprocessing2011})  -- see Table~\ref{tab:applications}.

The remainder of this paper is organized as follows: Section~\ref{sec:multimodels} gives a review of the dynamic borrowing methods we shall explore in this paper. 
In Section~\ref{sec:applications} we explore several applications of dynamic borrowing priors.
In Section~\ref{sec:survival} we discuss how to combine historical data, Bayesian analysis and leave-one-out cross-validation (LOO-CV) to improve inference for a piecewise exponential survival model.
Section~\ref{sec:log_reg} brings an application of Bayesian logistic regression for clinical trial data for which we also demonstrate how to use historical data to improve variable selection and average treatment effect estimation (Section \ref{sec:varsel}).
A crucial question is how different DBPs behave in terms of information use and shrinkage, which we tackle using real data in Section~\ref{sec:method_comparison}.
In Section~\ref{sec:propensity} we illustrate how to combine DBPs with propensity score matching in order to further control the amount of historical information borrowed.
Finally, Section~\ref{sec:discussion} brings a discussion of the results, the connections with the broader literature and directions for future research.

%% file: 002-models_hist_data.tex
\section{Models for incorporating historical data}
\label{sec:multimodels}

In this section we review four dynamic borrowing priors/methods, highlighting the commonalities and differences.
The methods vary in how they approach borrowing: from tempering the likelihood of the historical data directly \textit{via} a discounting factor as in the (normalized) power prior to leveraging the regularization properties of hierarchical models as with the robust meta-analytic prior to employing individual-level discounting as with the LEAP.
For ease of exposition, we shall focus on parametric regression models with regression coefficients $\bm{\beta}$ and dispersion parameter $\phi$ (e.g., a generalized linear model [GLM]), but the approaches described here are applicable much more broadly.
Figure~\ref{fig:borrowing_schematic} shows which parameters are shared between the current and historical data models under each prior, and Table~\ref{tab:methods_summary} summarizes the main features of the priors.

\input{fig_borrowing_schematic}
\input{tab_methods_summary}

\subsection{Normalized power prior (NPP)}
\label{sec:npp}

We begin our exposition with the power prior, which is a fixed borrowing method, in order to motivate the development of its \textit{normalized} (i.e., dynamic) version, which assigns a prior distribution to the discounting parameter in order to allow for dynamic borrowing.

\subsubsection{Power prior with fixed discounting parameter}

The power prior (PP) was proposed by \citet{Ibrahim2000} and involves discounting historical data by tempering the likelihood of the historical data by some value $a_{0} \in [0, 1]$ (often referred to as the \emph{discounting parameter}) along with eliciting an initial prior $\pi_0$.
We write
\begin{align}
    \pi_{\text{PP}}(\bm{\beta}, \phi | D_{0}, a_{0}, \pi_0) 
    = \frac{L(\bm{\beta}, \phi | D_{0})^{a_{0}} \pi_0(\bm{\beta}, \phi)}{Z(a_{0})}
    \propto L(\bm{\beta}, \phi | D_{0})^{a_{0}} \pi_0(\bm{\beta}, \phi),
    \label{eq:pp_fixeda0_singledataset}
\end{align}
where  $\bm{\beta}$ are regression coefficients, $\phi$ is a dispersion parameter, and $Z(a_{0}) = \int_{\mathbb{R}^p} \int_{0}^{\infty} L(\bm{\beta}, \phi | D_{0})^{a_{0}} \pi_0(\bm{\beta}, \phi) d\phi ~d\bm{\beta}$ is a normalizing constant which can be safely ignored when $a_{0}$ is fixed.
For fixed $a_{0}$, the \textbf{approximate} prior effective sample size~\citep{Morita2008,Wiesenfarth2020} of the PP is given by $a_{0} n_{0}$ -- see Section~\ref{sec:analysis_survival} for another formulation of the ESS.
Moreover, $a_{0} = 0$ sets the PP to the initial prior and $a_{0} = 1$ is equivalent to the regular sequential Bayesian update using the full posterior of the historical data as a prior to analyze the current data.
For multiple historical datasets the extension is straightforward:
\begin{align}
    \pi_{\text{PP}}(\bm{\beta}, \phi | D_0, \bm{a}_0, \pi_0) \propto \left[\prod_{h = 1}^{H} L(\bm{\beta}, \phi | D_{0h})^{a_{0h}}\right] \pi_0(\bm{\beta}, \phi),%
    \label{eq:pp_fixeda0}
\end{align}
and the initial prior $\pi_0$ is typically vague (non-informative), the rationale being that the historical data will bring the necessary information.
A common choice for regression problems is
\begin{align}
    \beta_j &\sim N(\mu_{0j},  \sigma_{0j}^2) \text{ for } j = 1, \ldots, p, \notag \\
    \phi &\sim N^{+}(\alpha_0, \gamma_0^2),
    \label{eq:pp_fixeda0_initialprior}
\end{align} 
where $N^{+}(\mu, \sigma^2)$ denotes the truncated normal distribution with location parameter $\mu$ and scale parameter $\sigma$ (i.e., the $N(\mu, \sigma^2)$ distribution truncated to the positive real line). Note that the half-normal distribution is the special case with $\mu = 0$.
The hyperparameters $\bm{\mu}_0 = (\mu_{01}, \ldots, \mu_{0p})'$, $\bm{\sigma}_0 = (\sigma_{01}, \ldots, \sigma_{0p})'$, $\alpha_0$, and $\gamma_0$ are elicited, but can be safely set at non-informative defaults.

The PP thus provides a flexible way to incorporate historical data and control the informativeness of the prior.
On the other hand, its informativeness depends crucially on $a_0$.
Section~5 of \citet{Ibrahim2015} gives an overview of how to select the
discounting parameter(s).

\subsubsection{Dynamic borrowing via the normalized power prior}

A common strategy to avoid having to select $\bm{a}_0$ is to take a fully Bayesian approach and place a prior $\pi_A$ on the discounting parameters.
In this setting, however, the normalizing constant must be included in the expression for the full posterior, otherwise the resulting posterior violates the likelihood principle \citep{Duan2006,Neuenschwander2009} and forces the marginal prior on $a_0$ to itself depend on the external data. This is problematic, because a prior for $a_0$ intended to be non-informative (e.g., uniform) may end up being informative.

To circumvent these limitations, the normalized power prior (NPP) was introduced. We write the NPP mathematically as
\begin{align}
    \pi_{\text{NPP}}(\bm{\beta}, \phi, \bm{a}_0 | D_0, \pi_0)
    &= \prod_{h=1}^H \pi_{\text{PP}}\left(\bm{\beta}, \phi | D_{0h}, a_{0h}, \pi_{0}^{1/H}\right)
    \pi_A(a_{0h})
    ,\notag \\
    &= \prod_{h=1}^H
    \frac{ L(\bm{\beta}, \phi | D_{0h})^{a_{0h}}
    \pi_0(\bm{\beta}, \phi)^{1/H}}
    {Z_h(a_{0h})} \pi_A(a_{0h})
    ,\notag\\
    &= \left[\prod_{h = 1}^{H} 
        \frac{L(\bm{\beta}, \phi | D_{0h})^{a_{0h}}
        }{Z_h(a_{0h})}
        \pi_A(a_{0h})
    \right]
    \pi_0(\bm{\beta}, \phi)
    \label{eq:npp}
\end{align}
where $Z_h(a_{0h}) = \int_{\mathbb{R}^p} \int_{0}^{\infty} L(\bm{\beta}, \phi | D_{0h})^{a_{0h}} \pi_0(\bm{\beta}, \phi)^{1/H} d\phi ~d\bm{\beta}$ is a normalizing constant and $\pi_A(a_{0h})$ is a prior on $a_{0h}$, usually taken to be a beta prior. The purpose of discounting the initial prior $\pi_0(\bm{\beta}, \phi)$ by a factor of $1/H$ is primarily computational: it allows the overall initial prior to be decomposed across the $H$ historical datasets while preserving $\prod_{h=1}^H \pi_0(\bm{\beta},\phi)^{1/H}=\pi_0(\bm{\beta},\phi)$. This decomposition facilitates the construction of each dataset-specific normalizing constant $Z_h(a_{0h})$ under a proper baseline density, thereby making the normalized power-prior formulation computationally tractable while avoiding the need to repeatedly normalize the full power likelihood jointly across all historical datasets.

In most settings, the function $Z_h(\cdot)$ is analytically intractable, and it must therefore be estimated numerically.
A common choice is to estimate $Z_h$ for a few values of $a_{0h}$ using an accurate algorithm such as bridge sampling~\citep{Meng1996}, which can be accomplished using the \textbf{bridgesampling} package in R \citep{Gronau2020} and then interpolate using a smooth -- possibly shape-constrained -- function.
We refer the reader to \citet{Carvalho2021} for more details on this two-step approach, which is implemented in the R packages \textbf{hdbayes}~\citep{Alt2025} and  \textbf{BayesPPD}~\citep{Shen2022}. 
For generalized linear models (GLMs) there are alternative approaches.
For instance, the R package \textbf{NPP}~\citep{NPP2023} offers an implementation of the NPP for GLMs which employs a Laplace approximation \citep{Tierney1986} to estimate the normalizing constant, which is typically much faster than the fully Bayesian implementation, but can be inaccurate for small samples and/or higher dimensional problems.

\subsection{Robust meta-analytic prior (rMAP)}
\label{sec:rmap}

One may also leverage the power of more traditional hierarchical modeling tools, in which the data-generating process is specified sequentially at different population levels.
In this context, we now turn our attention to the meta-analytic approach of \cite{Neuenschwander2010, spiegelhalter2004bayesian}, which relies on the Bayesian hierarchical model (BHM).
While here we focus on the parametric form of the model, non-parametric and semi-parametric approaches are also possible~\citep{Hupf2021}. 

The BHM assumes that the parameters for the current and historical data sets are different, but come from the same distribution, whose hyperparameters are then estimated.
Let $(\bm{\beta}', \phi)'$ denote the GLM parameters for the current data set, where $\bm{\beta} = (\beta_1, \ldots, \beta_p)'$. Let $(\bm{\beta}_{0h}', \phi_{0h})'$ denote the GLM parameters for the historical data set $D_{0h}$, $h = 1, \ldots, H$, where $\bm{\beta}_{0h} = (\beta_{0h1}, \ldots, \beta_{0hp})'$.
The BHM can be written as
\begin{align}
    \mu_j | \mu_{0j}, \sigma_{0j} &\sim N(\mu_{0j}, \sigma_{0j}^2)
        , \ \ j = 1, \ldots, p
    ,\notag \\
    \sigma_j | m_j, s_j &\sim N^+(m_j, s_j^2)
        , \ \ j = 1, \ldots, p
    ,\notag \\
    \beta_j, \beta_{0hj} | \mu_j, \sigma_j &\overset{\text{i.i.d.}}{\sim} N(\mu_j, \sigma_j^2)
        , \ \ h = 1, \ldots, H
        , \ \ j = 1, \ldots, p
    ,\notag \\
    \phi | m_0, s_0 &\sim N^{+}(m_0, s_0^2)
        , \notag \\
    \phi_{0h} | m_{0h}, s_{0h} &\sim N^{+}(m_{0h}, s_{0h}^2)
        , \ \  h = 1, \ldots, H
    ,\notag \\
    y_i    | \bm{\beta}, \phi &\sim f(y_i | \bm{\beta}, \phi)
        , \ \ i = 1, \ldots, n
    , \notag \\
    y_{0hi} | \bm{\beta}_{0h}, \phi_{0h} &\sim f(y_{0hi} | \bm{\beta}_{0h}, \phi_{0h}), 
        \ \ h = 1, \ldots, H, 
        \ \ i = 1, \ldots, n_{0h}, 
    \label{eq:bhm}
\end{align}
where $f(\cdot | \bm{\beta}, \phi)$ is the density (or mass) function corresponding to the GLM likelihood, $\mu_j$ is referred to as the global (or meta-analytic) mean,  $\sigma_j$ is the global standard deviation -- which reflects the heterogeneity between the data sets --  
and $\bm{\xi} = ( m_0, s_0, \{(m_{0h}, s_{0h}), h = 1, \ldots, H \}, \{ (\mu_{0j}, \sigma_{0j}, m_j, s_j) , j = 1, \ldots, p \} )$ are elicited hyperparameters.
The standard deviation hyperparameters $s_j$ are worth the most attention as they control most of the borrowing properties under the BHM.

Let $\bm{\theta}$ denote all parameters to be sampled as in \eqref{eq:bhm}.
The resulting posterior density may be expressed as
\begin{align}
    p_{\text{BHM}}(\bm{\theta} | D, D_0, \bm{\xi}) \propto
    L(\bm{\beta}, \phi | D)
    \left[
        \prod_{h=1}^H L(\bm{\beta}_{0h}, \phi_{0h} | D_{0h})
    \right]
    \pi_{\text{BHM}}(\bm{\theta})
    ,
    \label{eq:bhm_post}
\end{align}
where
\begin{align}
    \pi_{\text{BHM}}(\bm{\theta}) &=
    \varphi^+(\phi | m_0, s_0^2)
    \left[ \prod_{h=1}^H \varphi^+(\phi_{0h} | m_{0h}, s_{0h}^2) \right] \cdot \notag \\
    &~~~~~ \left\{
    \prod_{j=1}^p 
    \left[ \varphi(\mu_j | \mu_{0j}, \sigma_{0j}^2) \varphi^+(\sigma_j | m_j, s_j^2)
    \varphi(\beta_j | \mu_j, \sigma_j^2) \prod_{h=1}^H \varphi(\beta_{0hj} | \mu_j, \sigma_j^2)
    \right]
    \right\},
    \label{eq:bhm_prior}
\end{align}
with $\varphi(\cdot | a, b^2)$ and $\varphi^+(\cdot | a, b^2)$ denoting the densities of the normal distribution $N(a, b^2)$ and the truncated normal distribution $N^{+}(a, b^2)$, respectively. 

The meta-analytic predictive (MAP) prior \citep{Neuenschwander2010,spiegelhalter2004bayesian} is simply the prior for the current data parameters $(\bm{\beta}, \phi)$ induced by the BHM.
It is given a special name because, in a BHM, the meta-analytic mean or the individual groups (e.g., parameters for each subgroup) are typically of interest, whereas when borrowing information from historical data, only the current data parameters are of interest.
More fundamentally, the MAP formulation allows for the prior in an analysis to be completely pre-specified without observing the current data, which is useful for regulatory submissions.

The marginal MAP on the coefficients may be expressed mathematically as
\begin{align} 
    \pi_{\text{MAP}}(\bm{\beta} | D_0)
        &= \int \int \left[\prod_{j=1}^p \varphi(\beta_j | \mu_j, \sigma_j^2) \right] \pi(\bm{\mu}, \bm{\sigma} | D_0) d\bm{\mu} d\bm{\sigma},
        \label{eq:map}
\end{align}
where $\bm{\mu} = (\mu_1, \ldots, \mu_p)$, $\bm{\sigma} = (\sigma_1, \ldots, \sigma_p)$, and $\pi(\bm{\mu}, \bm{\sigma} | D_0)$ is the posterior density of $(\bm{\mu}, \bm{\sigma})$ obtained by a BHM using only the historical data, i.e.,
\begin{align}
    \pi(\bm{\mu}, \bm{\sigma} | D_0) &= 
    \int \int p_{\text{BHM}}(\bm{\mu}, \bm{\sigma}, \bm{\beta}_0, \bm{\phi}_0 | D_0) d\bm{\beta}_0 d\bm{\phi}_0
    ,\notag \\
    &\propto \int \int \left\{\prod_{h=1}^H L(\bm{\beta}_{0h}, \phi_{0h} | D_{0h}) \varphi^+(\phi_{0h} | m_{0h}, s_{0h}^2) \prod_{j=1}^p \varphi(\beta_{0hj} | \mu_j, \sigma_j^2)
    \right\}
    \notag\\
    &\qquad\qquad \cdot
    \left\{\prod_{j=1}^p
      \varphi(\mu_j | \mu_{0j}, \sigma_{0j}^2)\,
      \varphi^+(\sigma_j | m_j, s_j^2)
    \right\} d\bm{\beta}_0\, d\bm{\phi}_0,
    \label{eq:map_mean_sd}
\end{align}
where $p_{\text{BHM}}(\cdot | D_0)$ is the posterior density of the BHM for the historical data, which is computed similarly to \autoref{eq:bhm_post} without the factor pertaining to the historical data, $L(\bm{\beta}, \phi | D)$.
Moreover, $\bm{\beta}_0 = (\bm{\beta}_{01}', \ldots, \bm{\beta}_{0H}')'$, and $\bm{\phi}_0 = (\phi_{01}, \ldots, \phi_{0H})'$.

The \textbf{robust} meta-analytic predictive (rMAP) prior \citep{Schmidli2014} is thus a two-component mixture prior consisting of a meta-analytic predictive (MAP) prior using the historical data sets and a vague prior, with weight parameter $\gamma \in (0, 1)$.
Similarly to the power prior, if $\gamma = 0$, the rMAP prior is simply the vague prior, whereas when $\gamma = 1$, the rMAP is the MAP, which in turn reduces to the BHM with the current data as a further component.
The self-adapting mixture (SAM) approach of~\citet{Yang2023} sets the weight using a clinically relevant difference $\delta$ and the likelihood ratio of the current data under different allowable configurations under $\delta$.
The resulting prior therefore depends on the current outcomes, and the SAM approach, including its propensity-score-integrated~\citep{Zhao2025} and covariate-adjusted~\citep{Park2026} extensions, is an empirical Bayes procedure.
The same holds for the elastic prior~\citep{Jiang2023}, which inflates the variance of the informative prior through a function of a congruence statistic computed from the current and historical data.
In the priors considered here, the current data enter the analysis only through the likelihood, apart from the use of baseline covariates in the propensity score methods, so that the posterior is obtained by a standard Bayesian update and, for the dynamic priors, the degree of borrowing is estimated jointly with the other parameters; we thus restrict our comparisons to this class.
The PP, NPP, BHM, rMAP, CP and LEAP are, moreover, defined for the full parameter vector of a regression model, and can therefore be combined directly with covariate adjustment, variable selection and $g$-computation, as in Section~\ref{sec:applications}.
For the SAM, there is the further difficulty of setting a clinically relevant difference in such a way that it is comparable to the other priors discussed here.

For i.i.d. settings, one can ease the computational burden by approximating the MAP prior with a mixture of conjugate priors~\citep{Schmidli2014}.
This approach is implemented in the R package \textbf{RBesT}~\citep{Weber2021}.
In regression settings, although a conjugate prior exists \citep{chen_conjugate_2003}, it is not possible to sample from the distribution without MCMC.
It is, however, possible to approximate the MAP with, say, a finite mixture of normal distributions.
This involves (i) fitting the MAP with MCMC, (ii) approximating the MAP with a finite mixture model (e.g., via maximum likelihood treating the MCMC samples as data), and subsequently sampling from the posterior using the mixture approximation.
In high dimensions however, the mixture approximation can become fragile and the posterior fit can become computationally burdensome.

Alternatively, one may use the marginal likelihood of the non-informative and meta-analytic predictive priors.
Specifically, note that
\begin{align}
    p_{\text{rMAP}}(\bm{\beta}, \phi | D, D_0, \gamma) &= 
    \frac{
    L(\bm{\beta}, \phi | D) \left[ \gamma \pi_I(\bm{\beta}, \phi | D_0) + (1 - \gamma) \pi_V(\bm{\beta}, \phi) \right]
    }{
        \int \int
        L(\bm{\beta}^*, \phi^* | D) \left[ \gamma \pi_I( \bm{\beta}^*, \phi^* | D_0) + (1 - \gamma) \pi_V( \bm{\beta}^*, \phi^* ) \right] 
        d\bm{\beta}^*d\phi^*
    }
    ,
    \notag \\
    &= \tilde{\gamma} p_I(\bm{\beta}, \phi | D, D_0) + (1 - \tilde{\gamma}) p_V(\bm{\beta}, \phi | D),
\end{align}
where $p_I(\bm{\beta}, \phi | D, D_0) = L(\bm{\beta}, \phi | D) \pi_I(\bm{\beta}, \phi | D_0) / Z_I(D, D_0)$ is the posterior density under the informative prior, $p_V(\bm{\beta}, \phi | D) = L(\bm{\beta}, \phi | D) \pi_V(\bm{\beta}, \phi) / Z_V(D)$ is the posterior density under the vague prior, and
\begin{align}
    \tilde{\gamma} = \frac{ \gamma Z_I(D, D_0) }{ \gamma Z_I(D, D_0) + (1 - \gamma) Z_V(D) }
    \label{eq:rmap_weight}
\end{align}
is the updated mixture weight.
The normalizing constants $Z_I(D, D_0)$ and $Z_V(D)$ are estimated via \textbf{bridgesampling} \citep{Gronau2020} in \textbf{hdbayes}.

\subsection{Commensurate prior}
\label{sec:commensurate}

Similarly to the rMAP, the commensurate prior (CP, \cite{Hobbs2012}) is a hierarchical prior. 
In the traditional CP, the regression coefficients for the current data, $\bm{\beta} = (\beta_1, \ldots, \beta_p)'$ are assumed to have a normal distribution centered at the historical data regression coefficients, $\bm{\beta}_0 = (\beta_{01}, \ldots, \beta_{0p})'$, and a hierarchical precision parameter. 
Moreover, the CP assumes $\beta_j \sim N(\beta_{0j}, \tau_j^{-1})$, where $\tau_j$ is referred to as the ``commensurability parameter'' for regression coefficient $j$.
This parameter measures how compatible the current and historical data are based on the $j^{th}$ covariate, with higher values indicating a larger degree of commensurability.
However, the numerical scale of $\tau_j$ depends on the parameterization of the model and the corresponding covariate, making it difficult to interpret the magnitude of $\tau_j$ in general.

We employ the extension of the CP to multiple historical data sets proposed by~\citet{Alt2025}, where it is assumed that the historical data sets have the same regression coefficients but potentially different dispersion parameters.
This formulation of the CP -- as implemented in the \textbf{hdbayes} package -- is given by
\begin{align}
   \beta_{0j} | \mu_{0j}, \sigma_{0j} &\sim N(\mu_{0j}, \sigma_{0j}^2), \ \ j = 1, \ldots, p
   , \notag \\
   \phi | m_0, s_0 &\sim N^+(m_0, s_0^2)
   ,\notag \\
   \phi_{0h} | m_{0h}, s_{0h}  &\sim N^+(m_{0h}, s_{0h}^2), \ \ h = 1, \ldots, H 
   ,\notag \\
   \beta_j | \beta_{0j}, \tau_j &\sim N\left( \beta_{0j}, \tau_j^{-1} \right),
        \ \ j = 1, \ldots, p
    ,\notag \\
    \tau_j | p_{\text{spike}}, \mu_{\text{spike}}, \sigma_{\text{spike}}, \mu_{\text{slab}}, \sigma_{\text{slab}} &\sim 
    p_{\text{spike}} N^+(\mu_\text{spike}, \sigma_{\text{spike}}^2)
    + (1 - p_{\text{spike}}) N^+(\mu_{\text{slab}}, \sigma_{\text{slab}}^2)
    , \notag \\
    y_i    | \bm{\beta}, \phi &\sim f(y_i | \bm{\beta}, \phi), \ \  i = 1, \ldots, n
    , \notag \\
    y_{0hi} | \bm{\beta}_0, \phi_{0h} &\sim f(y_{0hi} | \bm{\beta}_0, \phi_{0h}), \ \ h = 1, \ldots, H, \ \  i = 1, \ldots, n_{0h}.
    \label{eq:comm_hm}
\end{align}
When there is only one historical data set the CP as formulated here will correspond to the original commensurate prior. 
The Stan implementation in \textbf{hdbayes} marginalizes over the discrete indicators in the spike-and-slab prior thus leading to a continuous approximation.

Since the $\tau_j$ are prior \textit{precisions}, large values concentrate $\beta_j$ near $\beta_{0j}$ and thus induce heavy borrowing, whereas values near zero leave $\beta_j$ nearly free of the historical estimate. 
One may place a spike-and-slab prior on the commensurability parameters~\citep{Hobbs2012,Ouma2022,Zheng2022,Alt2025,Nikolaidis2025}.
We set the hyperparameters such that the ``spike'' approximates a point mass at $\tau_j = 200$ and encourages a high degree of borrowing. 
Conversely, the default hyperparameters for the slab lead to a prior which is approximately uniform on $(0, 3)$ and whose tail decays rapidly for $\tau_j \gg 3$ thus encouraging little borrowing.
These are, however, general recommendations, which should be scrutinized for each problem.
Propensity-score-based implementations of the CP are available from the R packages \textbf{psborrow} and \textbf{psborrow2} \citep{psborrow_package, psborrow2_package}.
The package \textbf{psborrow} employs~\textbf{rjags}~\citep{Plummer2025} to sample from the resulting posterior \textit{via} MCMC
while \textbf{psborrow2} uses a modern implementation of Hamiltonian Monte Carlo~\citep{Duane1987,Neal2011,Betancourt2017} available in the R package \textbf{cmdstanr}~\citep{Gabry2025}. 

\subsection{Latent exchangeability prior (LEAP)}
\label{sec:leap}

The previous approaches to dynamic borrowing are similar in the sense that they all assume each subject in the external data is equally relevant.
In practice, some subjects in the external data set may be more similar to subjects in the current data set than others.
One recently developed approach that directly accommodates this is the
latent exchangeability prior (LEAP, \cite{altLEAP2024}).
The LEAP assumes that the historical data are generated from a finite mixture model consisting of $K \ge 2$ components, with the current data generated from the first component of the mixture.
This allows for dynamic borrowing at the individual level, while the previous priors implement blanket discounting of the external data.

For a single historical data set, the Bayesian model under the LEAP can be written as
\begin{align}
    \bm{\gamma} = (\gamma_1, \ldots, \gamma_K)' &\sim \text{Dirichlet}(\bm{\alpha}_0)
    ,\notag \\
    \bm{\beta}, \bm{\beta}_{0k} 
        &\overset{\text{i.i.d.}}{\sim}
        N(\bm{\mu}_0, \bm{\Sigma}_0)
    , \ \ k = 2, \ldots, K
    ,\notag \\
    \phi, \phi_{0k}
        &\overset{\text{i.i.d.}}{\sim}
        N^+(m_0, s_0^2)
    , \ \ k = 2, \ldots, K
    ,\notag \\
    y_i | \bm{\beta}, \phi &\sim f(\cdot | \bm{\beta}, \phi)
    , \ \  i = 1, \ldots, n
    , \notag \\
    y_{0i} | \bm{\beta}, \bm{\beta}_0, \phi, \bm{\phi}_0, \bm{\gamma} &\sim \gamma_1 f(\cdot | \bm{\beta}, \phi) + \sum_{k=2}^K \gamma_k f(\cdot | \bm{\beta}_{0k}, \phi_{0k}), 
     \ \ i = 1, \ldots, n_0,
    \label{eq:leap}
\end{align}
where $\bm{\gamma} = (\gamma_1, \ldots, \gamma_K)$ are the mixing probabilities, $\bm{\alpha}_0 = (\alpha_{01}, \ldots, \alpha_{0K})'$ is a concentration hyperparameter, $\bm{\mu}_0$ and $\bm{\Sigma}_0$ respectively are prior mean and covariance matrices for the $K$ regression coefficients, and $m_0$ and $s_0$ are respectively the location and scale parameter for the truncated normal prior on the $K$ dispersion parameters.
The default hyperparameters are chosen to be consistent with non-informative initial priors.

The parameter $\gamma_1$ is a \emph{global} discounting parameter that tempers the influence of the external data on the resulting posterior density. The approach implicitly conducts \emph{local} discounting at the individual external trial participant level. As discussed by \citet{altLEAP2024}, the LEAP has the attractive interpretation as a prior over all partitions of the external data into exchangeable and non-exchangeable groups. It is mathematically equivalent to conducting Bayesian model averaging (BMA) over all such partitions.
Multisource exchangeability models~\citep{Kaizer2018} also average over configurations of exchangeability, but at the level of data sources rather than individuals.
The LEAP is implemented in the R package \textbf{hdbayes}.

\subsection{``Integrating'' data-based priors with the propensity score}\label{sec:ps}
Thus far, we have discussed several information borrowing priors and more or less assumed that the current and external data sets were sufficiently similar. In practice, this is not likely to be plausible. While the dynamic borrowing approaches provide some safeguarding against heterogeneity, in general, it is advisable to incorporate any \emph{prior} information about which subset of the external data is more likely to be exchangeable.

Recently, there have been many contributions to the literature surrounding so-called ``propensity-score integrated'' priors (see, e.g., \cite{wangPSIPP2019,liu2021propensity,li2022augmenting,lin2022incorporating}).
Most of these priors consist of a design stage in which the propensity score (PS) is estimated followed by an analysis stage in which the analysis is conducted (incorporating the PS-adjusted external data).
Recent examples include the propensity score weighted power prior~\citep{Polak2025}, the matching-assisted power prior~\citep{Qian2025} and propensity score weighted multisource exchangeability models~\citep{Wei2024}.
Although the PS integrated priors are sophisticated, it is straightforward to adapt any of the aforementioned priors while incorporating the propensity score.
Perhaps the easiest approach is to use propensity score matching (PSM, \cite{caliendo2008some}), which we briefly review.

Consider a data set in which we stack the current and external data sets together, creating the variable $S \in \{0, 1\}$, where $S = 1$ refers to data from the current study and $S = 0$ refers to data from the external study.
Let $N$ refer to the total number of subjects from all data sets.
We can do the following to include subjects that most closely match the design of the current data in the Bayesian analysis:
\begin{enumerate}
    \item Estimate $e_i \equiv \widehat{\Pr}\left( S_i = 1 | \bm{X} = \bm{x}_i \right), i = 1, \ldots, N$ (e.g., through logistic regression or some nonparametric regression techniques).
    \item Use a $k$-nearest neighbors (kNN) algorithm to find the $k$ subjects from the external data with propensity scores closest to the current data.
    \item Keep \emph{all} current data subjects and \emph{matched} external data subjects and proceed with the Bayesian analysis.
\end{enumerate}

In an external control arm augmentation problem, typically, $k = 1$ in step (2) above so that we find the external control subjects with baseline covariates as close as possible to the current data subjects.
If the amount of subjects to be borrowed must exceed the current study data, we may choose $k > 1$ so that multiple external data subjects are matched to a single current data subject.

Since PSM effectively discards a subset of the external data, the reader may ask: what is the point of doing PSM when the dynamic borrowing priors already discount the external data?
One reason is the bias-variance tradeoff.
In practice, it is rare that the current and external populations are the same.
In these cases, the dynamic borrowing priors may discount too little, increasing bias.
Conversely, they may discount too much, negating any efficiency gains by borrowing.
For example, an interesting simulation exercise showed that, when the number of exchangeable external data subjects is fixed but the total number of external data observations increases, any efficiency gains from the LEAP vanish \citep{campbell2024discussion}.
Recent work has shown through simulation studies that combining dynamic borrowing (e.g., commensurate prior) with PSM can help reduce bias and improve coverage probabilities toward the nominal level when discrepancies exist between current and external data \citep{yamaguchiHybridControlDesign}.
\citet{Jesse2026} give a neutral comparison of the propensity score weighted commensurate prior, propensity score weighted multisource exchangeability models and Bayesian additive regression trees for borrowing historical control data.
Since the PSM approach is easy to implement for any of the aforementioned priors, it can additionally be used as a sensitivity analysis.

An equally valid question is: if PSM itself will make the historical data more relevant, why add on the DBP layer?
The main reason is that PSM merely accounts for discrepancies between \emph{baseline} characteristics.
Borrowing external data conditional on baseline covariates requires exchangeability assumptions similar to those used to extend inferences from a trial to a new target population~\citep{Dahabreh2020}.
Since the external data is typically older, there could be outcome drift that would not be captured by propensity score approaches.
In external control arm augmentation, for example, the control arm is typically a standard of care (SOC), and the skill with which clinicians treat patients with the SOC may improve over time.
Another reason is that regulators may limit the amount of information (i.e., effective sample size) that can be borrowed, requiring a discounting of the matched data even if the populations are genuinely exchangeable.

%% file: fig_borrowing_schematic.tex
\providecommand{\aifignote}{}
\begin{figure}[tbp]
\centering
\tikzset{
  par/.style={circle, draw, minimum size=9mm, inner sep=0pt, font=\footnotesize},
  obs/.style={par, fill=gray!20},
  fixed/.style={rectangle, draw, minimum size=7mm, inner sep=1pt, font=\footnotesize},
  arr/.style={-{Latex[length=1.8mm]}},
  plate/.style={draw, rounded corners=2pt, inner sep=4pt},
  platelab/.style={font=\scriptsize, anchor=north east, inner sep=1.5pt},
  note/.style={font=\scriptsize, align=center},
  elab/.style={font=\scriptsize, midway},
}
\begin{subfigure}[t]{0.49\textwidth}
\centering
\begin{tikzpicture}
  \node[par] (theta) at (0, 0) {$\bm{\theta}$};
  \node[obs] (D) at (-1.4, -1.8) {$D$};
  \node[obs] (D0) at (1.4, -1.8) {$D_{0h}$};
  \node[par] (a0) at (2.9, -0.6) {$a_{0h}$};
  \draw[arr] (theta) -- (D);
  \draw[arr] (theta) -- (D0);
  \coordinate (m) at ($(theta)!0.5!(D0)$);
  \draw[arr, dashed] (a0) -- (m);
  \node[plate, fit=(D0)(a0)(m)] (pl) {};
  \node[platelab] at (pl.south east) {$h = 1, \ldots, H$};
  \node[note] at (0.4, -3.5) {$\pi(\bm{\theta}) \propto \prod_{h} L(\bm{\theta} \mid D_{0h})^{a_{0h}}\, \pi_0(\bm{\theta})$\\ $a_{0h}$ fixed (PP) or with prior $\pi_A$ (NPP)};
\end{tikzpicture}
\caption{PP and NPP}
\end{subfigure}\hfill
\begin{subfigure}[t]{0.49\textwidth}
\centering
\begin{tikzpicture}
  \node[par, minimum size=11mm] (mu) at (0, 0) {$\mu_j, \sigma_j$};
  \node[par] (b) at (-1.4, -1.6) {$\beta_j$};
  \node[par] (b0) at (1.4, -1.6) {$\beta_{0hj}$};
  \node[obs] (D) at (-1.4, -3.0) {$D$};
  \node[obs] (D0) at (1.4, -3.0) {$D_{0h}$};
  \node[fixed] (pv) at (-3.0, -1.6) {$\pi_V$};
  \draw[arr] (mu) -- node[elab, above left] {$\gamma$} (b);
  \draw[arr] (mu) -- (b0);
  \draw[arr, dashed] (pv) -- node[elab, above] {$1-\gamma$} (b);
  \draw[arr] (b) -- (D);
  \draw[arr] (b0) -- (D0);
  \node[plate, fit=(b0)(D0)] (pl) {};
  \node[platelab] at (pl.south east) {$h = 1, \ldots, H$};
  \node[note] at (-0.4, -4.4) {BHM and MAP: $\gamma = 1$; rMAP: $0 < \gamma < 1$};
\end{tikzpicture}
\caption{BHM, MAP and rMAP}
\end{subfigure}

\vspace{4mm}

\begin{subfigure}[t]{0.49\textwidth}
\centering
\begin{tikzpicture}
  \node[par] (tau) at (-1.4, 0) {$\tau_j$};
  \node[par] (b0) at (1.4, 0) {$\beta_{0j}$};
  \node[par] (b) at (-1.4, -1.6) {$\beta_j$};
  \node[obs] (D) at (-1.4, -3.0) {$D$};
  \node[obs] (D0) at (1.4, -3.0) {$D_{0h}$};
  \draw[arr] (tau) -- (b);
  \draw[arr] (b0) -- (b);
  \draw[arr] (b) -- (D);
  \draw[arr] (b0) -- (D0);
  \node[plate, fit=(D0)] (pl) {};
  \node[platelab] at (pl.south east) {$h = 1, \ldots, H$};
  \node[note] at (0, -4.4) {$\beta_j \sim N(\beta_{0j}, \tau_j^{-1})$; spike-and-slab prior on $\tau_j$};
\end{tikzpicture}
\caption{CP}
\end{subfigure}\hfill
\begin{subfigure}[t]{0.49\textwidth}
\centering
\begin{tikzpicture}
  \node[par] (theta) at (-1.6, -1.4) {$\bm{\theta}$};
  \node[obs] (D) at (-1.6, -3.0) {$D$};
  \node[par] (g) at (0.6, 0) {$\bm{\gamma}$};
  \node[par] (c) at (0.6, -1.4) {$c_i$};
  \node[obs] (y0) at (0.6, -3.0) {$y_{0i}$};
  \node[par] (tk) at (2.6, -1.4) {$\bm{\theta}_{0k}$};
  \draw[arr] (theta) -- (D);
  \draw[arr] (g) -- (c);
  \draw[arr] (c) -- (y0);
  \draw[arr] (theta) -- (y0);
  \draw[arr] (tk) -- (y0);
  \node[plate, fit=(c)(y0)] (pli) {};
  \node[platelab] at (pli.south east) {$i = 1, \ldots, n_0$};
  \node[plate, fit=(tk)] (plk) {};
  \node[platelab, anchor=south east] at (plk.north east) {$k = 2, \ldots, K$};
  \node[note] at (0.4, -4.4) {$c_i = 1$: subject $i$ shares $\bm{\theta}$ with the current data};
\end{tikzpicture}
\caption{LEAP}
\end{subfigure}
\caption{\textbf{Borrowing structure of the priors in Section~\ref{sec:multimodels}.}
Circles denote unknown parameters, shaded circles observed data, the square a fixed prior density, and rounded rectangles replication over historical data sets ($h$), mixture components ($k$) or historical subjects ($i$).
Dispersion parameters are omitted, and for the BHM and CP a single coefficient $j$ is shown.
(a) The current and historical data share all parameters $\bm{\theta} = (\bm{\beta}, \phi)$, and the historical likelihood is raised to the power $a_{0h}$ (dashed arrow).
(b) The current and historical coefficients are drawn from a common distribution with mean $\mu_j$ and standard deviation $\sigma_j$; the rMAP prior for $\beta_j$ is a mixture of the MAP prior induced by this hierarchy (weight $\gamma$) and a vague prior $\pi_V$ (weight $1 - \gamma$).
(c) The current coefficient is centered at a historical coefficient common to all historical data sets, with precision $\tau_j$.
(d) Each historical subject belongs to one of $K$ mixture components, with latent indicator $c_i$ and mixing probabilities $\bm{\gamma}$; only subjects in the first component share the parameters of the current data.\aifignote}
\label{fig:borrowing_schematic}
\end{figure}

%% file: tab_methods_summary.tex
\begin{table}[tbp]
\centering
\caption{\textbf{Summary of the borrowing priors.} ``Discounting unit'' is the level at which the influence of the external data is reduced.
All priors except the PSIPP and PSM depend only on the historical data and on hyperparameters; the PSIPP and PSM also use the baseline covariates of the current study, but not its outcomes.}
\label{tab:methods_summary}
\footnotesize
\renewcommand{\arraystretch}{1.25}
\begin{tabularx}{\textwidth}{>{\raggedright\arraybackslash}p{1.6cm} >{\raggedright\arraybackslash}X >{\raggedright\arraybackslash}p{1.7cm} >{\raggedright\arraybackslash}X >{\raggedright\arraybackslash}X >{\raggedright\arraybackslash}X}
\toprule
Prior & Mechanism & Discounting unit & Borrowing governed by & Normalizing constant & Several historical data sets \\
\midrule
PP (\S\ref{sec:npp}) & Historical likelihood raised to a fixed power & Data set & $a_{0h}$, fixed & Not needed & One $a_{0h}$ per data set, Eq.~\eqref{eq:pp_fixeda0} \\
NPP (\S\ref{sec:npp}) & As the PP, with a prior on the power & Data set & Prior $\pi_A$ on $a_{0h}$ & $Z_h(a_{0h})$, estimated by bridge sampling on a grid and interpolated & One $a_{0h}$ and $Z_h$ per data set, Eq.~\eqref{eq:npp} \\
BHM, MAP (\S\ref{sec:rmap}) & Exchangeable study-level coefficients, shrunk towards a common mean & Coefficient & Between-study SD $\sigma_j$ and its prior scale $s_j$ & Not needed & Each data set is a further study in the hierarchy \\
rMAP (\S\ref{sec:rmap}) & Mixture of the MAP prior and a vague prior & All historical data (MAP component) & Prior weight $\gamma$, updated to $\tilde{\gamma}$, Eq.~\eqref{eq:rmap_weight}; $s_j$ & $Z_I(D, D_0)$ and $Z_V(D)$, for $\tilde{\gamma}$ & Through the MAP prior \\
CP (\S\ref{sec:commensurate}) & Current coefficients centered at historical coefficients & Coefficient & Precision $\tau_j$ and its spike-and-slab hyperparameters & Not needed & Common $\bm{\beta}_0$, data-set-specific dispersion~\citep{Alt2025} \\
LEAP (\S\ref{sec:leap}) & Historical data modeled as a mixture, one component shared with the current data & Subject & Mixing probability $\gamma_1$ (Dirichlet prior) and number of components $K$ & Not needed & Data sets stacked into one~\citep{Alt2025} \\
PSIPP (\S\ref{sec:ps}, \S\ref{sec:survival}) & PS strata, PP within each stratum & Stratum & Stratum-specific $a_{0,k}$, set from the overlap of PS distributions & Not needed ($a_{0,k}$ fixed) & Not considered here \\
PSM with any prior (\S\ref{sec:ps}, \S\ref{sec:propensity}) & External subjects selected by PS matching before borrowing & Subject (selection), then as for the chosen prior & Caliper and matching ratio, then as for the chosen prior & As for the chosen prior & As for the chosen prior \\
\bottomrule
\end{tabularx}
\end{table}

%% file: 003-applications.tex
\section{Applications}
\label{sec:applications}

We now move on to consider two concrete applications of the DBPs discussed above in the broader context of Bayesian analysis of clinical trial data. Our goal is to showcase how external data can be incorporated and help improve the efficiency in standard statistical tasks such as variable selection and causal effect estimation. We begin with a survival analysis example, demonstrating the use of various DBPs in time-to-event models, and then proceed to a logistic regression example involving variable selection and Bayesian model averaging for estimating the average treatment effect (ATE).
Table~\ref{tab:applications} summarizes the two applications, and Figure~\ref{fig:workflow} lists the steps of the analysis together with the sections in which each step is illustrated.

\input{tab_applications}
\input{fig_workflow}

Most of the methods presented here rely on Markov chain Monte Carlo (MCMC) \textit{via} dynamic Hamiltonian Monte Carlo (dHMC, \cite{Betancourt2017}) implemented in the popular language Stan~\citep{Carpenter2017}.
Unless otherwise stated, we have employed the Stan defaults of four independent chains, each with $1000$ iterations for warmup (adaptation) and $1000$ iterations for sampling, with only the latter being retained for inference.
For all analyses we verified that no divergent transitions occurred during sampling and that the Monte Carlo standard error (MCSE) did not exceed 5\% of the marginal posterior standard deviation for all quantities.
Moreover, we have only retained runs for which the potential scale reduction factor (PSRF, $\hat{R}$) was below $1.05$~\citep{Vehtari2021} for all parameters.

\input{003_1-survival}
\input{003_2-logistic}
\input{003_3-method_comp}

%% file: tab_applications.tex
\begin{table}[tbp]
\centering
\caption{\textbf{Summary of the two applications.}}
\label{tab:applications}
\footnotesize
\renewcommand{\arraystretch}{1.25}
\begin{tabularx}{\textwidth}{>{\raggedright\arraybackslash}p{2.3cm} >{\raggedright\arraybackslash}X >{\raggedright\arraybackslash}X}
\toprule
 & Survival analysis (Section~\ref{sec:survival}) & Logistic regression (Sections~\ref{sec:log_reg}--\ref{sec:propensity}) \\
\midrule
Current study & E1690~\citep{E1690}; $n = 426$, 240 events & ACTG036~\citep{Merigan1991}; $n = 183$ \\
External study & E1684~\citep{E1684}; $n_0 = 262$, 175 events & ACTG019~\citep{Volberding1990}; $n_0 = 822$ ($103$ after PSM) \\
Treatment & High-dose interferon alfa-2b vs.\ observation & Zidovudine (AZT) vs.\ placebo \\
Outcome & Relapse-free survival & Death, AIDS or ARC (binary) \\
Outcome model & PWEPH and CurePWEPH, fitted separately by arm; $J$ selected by $\text{elpd}_{\text{loo}}$ & Logistic regression; covariate subsets combined by BMA \\
Covariates & Gender, age (standardized), multiple lymph node involvement & Age, race, treatment, \texttt{cd4} (age and \texttt{cd4} standardized) \\
Estimand & Difference in marginal two-year RFS probabilities, IFN minus control ($\Delta$) & Marginal risk difference (ATE); conditional odds ratios in Section~\ref{sec:method_comparison} \\
Priors & Vague (reference), PP ($a_0 = 0.5$), NPP, BHM, CP, LEAP, PSIPP & NPP (Sections~\ref{sec:varsel} and~\ref{sec:propensity}); NPP, BHM, rMAP, CP and LEAP, with pooling and no borrowing (Section~\ref{sec:method_comparison}) \\
Further analyses & Effective sample size; selection of model--prior pairs by $\text{elpd}_{\text{loo}}$ & Prior sensitivity analysis, prior predictive checks and pROC curves; PSM \\
\bottomrule
\end{tabularx}
\end{table}

%% file: fig_workflow.tex
\providecommand{\aifignote}{}
\begin{figure}[tbp]
\centering
\tikzset{
  wstep/.style={rectangle, draw, rounded corners=2pt, text width=7.6cm, align=left,
               inner sep=4pt, font=\footnotesize},
  woptstep/.style={wstep, dashed},
  wref/.style={font=\footnotesize, text width=4.6cm, align=left, anchor=west},
  arr/.style={-{Latex[length=1.8mm]}},
}
\begin{tikzpicture}[node distance=3.5mm]
  \node[wstep] (s1) {\textbf{1. Compare the data sources.} Baseline covariates, outcomes (e.g.\ Kaplan--Meier curves) and overlap of propensity scores.};
  \node[woptstep, below=of s1] (s2) {\textbf{2. Restrict or stratify the external data (optional).} Propensity score matching or stratification.};
  \node[wstep, below=of s2] (s3) {\textbf{3. Specify the outcome model and the borrowing prior(s).} Choose the prior(s) and elicit hyperparameters, including the prior on the borrowing parameter.};
  \node[wstep, below=of s3] (s4) {\textbf{4. Check the prior.} Prior predictive distributions and prior ROC curves.};
  \node[wstep, below=of s4] (s5) {\textbf{5. Fit the model and check convergence.} MCMC; divergent transitions, $\hat{R}$, Monte Carlo standard errors.};
  \node[wstep, below=of s5] (s6) {\textbf{6. Compare or average models.} $\text{elpd}_{\text{loo}}$, marginal likelihoods, Bayesian model averaging.};
  \node[wstep, below=of s6] (s7) {\textbf{7. Quantify the borrowing.} Effective sample size; posterior distribution of the borrowing parameter.};
  \node[wstep, below=of s7] (s8) {\textbf{8. Estimate marginal treatment effects.} Bayesian bootstrap over the covariate distribution ($g$-computation).};
  \node[wstep, below=of s8] (s9) {\textbf{9. Assess sensitivity.} Vary the initial prior, the prior on the borrowing parameter and the borrowing prior itself.};
  \foreach \a/\b in {s1/s2, s2/s3, s3/s4, s4/s5, s5/s6, s6/s7, s7/s8, s8/s9}
    \draw[arr] (\a) -- (\b);
  \node[wref] at ($(s1.east) + (3mm, 0)$) {Figure~\ref{fig:km_plots}; \S\ref{sec:survival}, \S\ref{sec:propensity}};
  \node[wref] at ($(s2.east) + (3mm, 0)$) {\S\ref{sec:ps}, \S\ref{sec:survival} (PSIPP), \S\ref{sec:propensity}; \textbf{psrwe}, \textbf{dbarts}, \textbf{MatchIt}};
  \node[wref] at ($(s3.east) + (3mm, 0)$) {\S\ref{sec:multimodels}; Tables~\ref{tab:supp_PWE_summary}, \ref{tab:supp_CurePWE_summary}; \textbf{hdbayes}};
  \node[wref] at ($(s4.east) + (3mm, 0)$) {\S\ref{sec:PSA}};
  \node[wref] at ($(s5.east) + (3mm, 0)$) {\S\ref{sec:applications}; Stan};
  \node[wref] at ($(s6.east) + (3mm, 0)$) {\S\ref{sec:survival}, \S\ref{sec:varsel}; \textbf{loo}, \textbf{bridgesampling}};
  \node[wref] at ($(s7.east) + (3mm, 0)$) {\S\ref{sec:analysis_survival}; Figure~\ref{fig:post_a0}};
  \node[wref] at ($(s8.east) + (3mm, 0)$) {\S\ref{sec:survival}, \S\ref{sec:ate}};
  \node[wref] at ($(s9.east) + (3mm, 0)$) {\S\ref{sec:PSA}, \S\ref{sec:method_comparison}, \S\ref{sec:propensity}};
\end{tikzpicture}
\caption{\textbf{Steps of an analysis with dynamic borrowing priors, as illustrated in Section~\ref{sec:applications}.} The right-hand column gives the sections, figures and tables in which each step is illustrated, and the R packages or software used. Step 2 (dashed) is optional.\aifignote}
\label{fig:workflow}
\end{figure}

%% file: 003_1-survival.tex
\subsection{Survival analysis using piecewise exponential and mixture cure rate models}
\label{sec:survival}

Informative priors are particularly valuable in time-to-event (TTE) analysis. In studies with TTE outcomes, the amount of information is based not on the total sample size, but on the number of observed events (i.e., failures).
When data are collected prospectively, this can prolong study durations and increase costs to infeasible levels.
In this section, we discuss how to incorporate external data into an analysis of a TTE trial using both a proportional hazards (PH) model with piecewise constant baseline hazards and a cure rate model.

\begin{figure}
    \centering
    \includegraphics[width=0.8\textwidth]{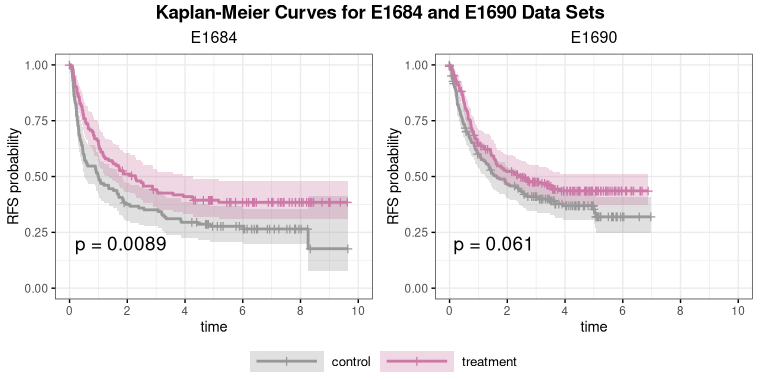}
    \caption{\textbf{Kaplan--Meier Curves for control and treatment arms in E1684 and E1690 studies}.}
    \label{fig:km_plots}
\end{figure}

We perform a Bayesian analysis of the E1690 study~\citep{E1690}, treated as the current data, with $n = 426$ subjects and 240 events, while incorporating external information from the E1684 study~\citep{E1684}, with $n_0 = 262$ subjects and 175 events.
Both studies were conducted by the Eastern Cooperative Oncology Group (ECOG) in high-risk melanoma patients and evaluated the effect of high-dose interferon alfa-2b (IFN) versus observation (control) on relapse-free survival (RFS).
Figure \ref{fig:km_plots} displays the Kaplan--Meier (KM) curves by treatment arm for each study. Compared with E1690, E1684 shows a more pronounced treatment effect and longer follow-up, suggesting potential incompatibility between the two data sets. Nevertheless, external information, if appropriately down-weighted or selectively incorporated, may still improve inference for the current trial. 
Hence, methods such as DBPs or propensity score-based approaches are particularly well suited in this setting.

Following previous analyses of these data sets \citep{chenBayesianCureRate2002, ibrahimBayesianMethodsClinical2012}, we consider two survival models: (1) a PH model with a piecewise constant baseline hazard (referred to as the piecewise exponential PH (PWEPH) model), and (2) a mixture cure rate model in which a fraction $\pi$ of the population is assumed to be cured, i.e., not at risk of the event during the study follow-up period.
The conditional survival function under the mixture cure rate model takes the form
$$
    S(t \mid \bm{X}) = \pi + (1 - \pi) S_{\text{PWEPH}}(t \mid \bm{X}),
$$
where $\bm{X}$ denotes the vector of baseline covariates, and $S_{\text{PWEPH}}(t \mid \bm{X})$ denotes the conditional survival function for the non-cured population, modeled using a PWEPH structure. Covariate effects are incorporated through the PWEPH component only, and no separate model is specified for the cure fraction $\pi$. We refer to this mixture cure rate model as the CurePWEPH model. We fit both models separately to each treatment arm. Both models can be implemented using the \textbf{hdbayes} package. Alternatively, a Bayesian analysis using the PWEPH model can also be carried out in \textbf{SAS} via \texttt{PROC PHREG}, although this procedure does not support cure rate models.
Prior work by \cite{chenBayesianCureRate2002} and \cite{ibrahimBayesianMethodsClinical2012} considered a promotion time cure rate model, which can be equivalently represented as a mixture cure rate model.
However, to our knowledge, there is no publicly available software that implements this approach with Bayesian borrowing. 

In E1690 and E1684 data sets, a small number of relapse times are recorded as 0 and are recoded as 0.5 days for analysis. For both the PWEPH model and the PWEPH component of the CurePWEPH model, we adjust for baseline covariates including gender, standardized age, and an indicator for multiple involved lymph nodes. The piecewise constant baseline hazards are constructed by partitioning the time axis into $J$ intervals with approximately equal numbers of events in the current data. We consider $J = 2, \ldots, 9$ and, separately for each treatment arm, select the value that maximizes the expected log predictive density (elpd), a measure of predictive accuracy computed using the Bayesian LOO-CV method. Intuitively, the LOO-CV estimate of elpd ($\text{elpd}_{\text{loo}}$) evaluates how well a model predicts each observation when that observation is excluded from model fitting. Higher elpd values indicate better predictive performance. In practice, we can approximate $\text{elpd}_{\text{loo}}$ using Pareto-smoothed importance sampling (PSIS; \citet{vehtari2017, Vehtari2024}) via the $\textbf{loo}$ R package \citep{loo_package}.

For prior specification,  we apply the informative priors described in Section \ref{sec:multimodels}, PP, NPP, BHM, CP, and LEAP, along with a vague (non-informative) prior. We also consider the propensity score-integrated power prior (PSIPP, \cite{wangPSIPP2019, luPropensityScoreintegratedPower2022}).
Full prior specifications for all methods under both the PWEPH and CurePWEPH models are provided in the Supplementary Material.
For the PP, we fix the discounting parameter at $a_0 = 0.5$ to down-weight the external data contribution.
Since LEAP has not been fully developed for mixture cure rate models, borrowing under LEAP is restricted to the PWEPH component for the non-cured population. The cure fraction is estimated solely from the current data and its initial prior, without borrowing historical information. A key motivation for this restriction is that the cure fraction is not identifiable under full borrowing via LEAP since a mixture of Bernoulli distributions is not identifiable without added structure. Under this construction, the PWEPH component in the external data model is represented as a two-component mixture. One component is identical to the PWEPH structure in the current data model, while the other allows for non-exchangeability through a distinct set of PWEPH model parameters. The external data model also includes its own cure probability.

For the PSIPP, we first estimate propensity scores (PS) by fitting a logistic regression model to the pooled current and external data sets, with study membership as the outcome (1 = current study, 0 = external study). The model includes an intercept and the same baseline covariates used in the survival models (i.e., PWEPH or CurePWEPH) under other priors.
All subjects are then stratified into four strata defined by the quartiles of the PS among the current-study subjects, so that each stratum contains approximately one quarter of these subjects; external subjects with PS outside the range observed in the current study are excluded.
Within each stratum $k \in \{1, \ldots, 4\}$, we apply the PP with a stratum-specific discounting parameter $a_{0, k}$, determined by the degree of overlap between the PS distributions of the two studies using the \textbf{psrwe} package \citep{psrwe_package} in R. Because covariate balance is achieved through stratification, a simplified survival model without baseline covariates is fitted within each stratum. External subjects more comparable to those in the current study (based on PS values) are assigned higher values of the discounting parameter $a_{0, k}$, resulting in less discounting of information from these strata.

In melanoma clinical trials, the two-year RFS probability is a commonly used threshold and milestone \citep{Melanoma13, Melanoma20}. To assess the effect of the high-dose IFN relative to the control, we define the treatment effect $\Delta$ as the difference in \textit{marginal} two-year RFS probabilities between subjects receiving high-dose IFN and those in the control group. To estimate these marginal probabilities, we use the Bayesian bootstrap approach \citep{rubinBayesianBootstrap1981}. For each posterior draw of the survival model parameters, we compute subject-specific two-year RFS probabilities under each treatment arm and average them over the empirical distribution of the covariates using Dirichlet weights. Repeating this procedure across posterior draws yields the posterior distribution of $\Delta$.
For the PSIPP, a similar procedure is applied within each PS stratum, with marginal RFS probabilities obtained by averaging stratum-specific estimates. A detailed description of the algorithm is included in the Supplementary Material.

\subsubsection{Analysis results}
\label{sec:analysis_survival}

For each treatment arm, we fit the CurePWEPH and PWEPH models across $J = 2, \ldots, 9$ intervals under each prior specification and compute the corresponding elpd values (shown in the Supplementary Material). For each combination of prior and model type (CurePWEPH or PWEPH), we then select the value of $J$ that maximizes elpd within each treatment arm for subsequent inference.
Posterior summaries of the estimated two-year RFS probabilities and the corresponding treatment effect $\Delta = \hat{S}_1(2) - \hat{S}_0(2)$ under the selected model-prior pairs are reported in Table \ref{tab:post_summ}. The estimated treatment effect varies across priors and model types. For example, the PWEPH model with the PSIPP yields the largest estimated treatment effect, with a posterior mean of $\Delta \approx 0.084$, posterior probability $p(\Delta > 0 \mid D) \approx 0.989$, and a $95\%$ Bayesian credible interval (BCI) excluding zero. In contrast, the PWEPH model with the vague prior produces the smallest estimated treatment effect (posterior mean $\approx 0.053$), with weaker evidence for benefit ($p(\Delta > 0 \mid D) \approx 0.868$) and a $95\%$ BCI that includes zero. Overall, LEAP, NPP, PP, and PSIPP under both model types tend to yield larger posterior means of $\Delta$, smaller posterior variance, and stronger evidence for a treatment benefit (i.e., higher $p(\Delta > 0 \mid D)$) than the other priors, suggesting that these approaches borrow more information from the external data.  

\begin{table}

\caption{Posterior Summaries of $\Delta$ under PWEPH and CurePWEPH Models}
\centering
\resizebox{\textwidth}{!}{
\begin{tabular}[t]{llcccccccc}
\toprule
\multicolumn{2}{c}{ } & \multicolumn{2}{c}{IFN} & \multicolumn{2}{c}{Control} & \multicolumn{4}{c}{$\Delta = \text{IFN - Control}$} \\
\cmidrule(l{3pt}r{3pt}){3-4} \cmidrule(l{3pt}r{3pt}){5-6} \cmidrule(l{3pt}r{3pt}){7-10}
Model & Prior & Mean & SD & Mean & SD & Mean & SD & $p(\Delta > 0 \mid D)$ & $95\%$ BCI\\
\midrule
CurePWEPH & Vague & 0.527 & 0.033 & 0.470 & 0.034 & 0.057 & 0.047 & 0.882 & $(-0.036, ~0.149)$\\
CurePWEPH & BHM & 0.525 & 0.033 & 0.466 & 0.034 & 0.059 & 0.047 & 0.896 & $(-0.033, ~0.150)$\\
CurePWEPH & CP & 0.526 & 0.033 & 0.468 & 0.034 & 0.057 & 0.047 & 0.884 & $(-0.035, ~0.151)$\\
CurePWEPH & LEAP & 0.533 & 0.031 & 0.463 & 0.031 & 0.070 & 0.044 & 0.945 & $(-0.016, ~0.157)$\\
CurePWEPH & NPP & 0.521 & 0.027 & 0.448 & 0.028 & 0.073 & 0.039 & 0.969 & $(-0.003, ~0.149)$\\
CurePWEPH & PP~$(a_0 = 0.5)$ & 0.523 & 0.029 & 0.452 & 0.029 & 0.071 & 0.041 & 0.958 & $(-0.009, ~0.151)$\\
CurePWEPH & PSIPP & 0.525 & 0.026 & 0.461 & 0.025 & 0.064 & 0.037 & 0.959 & $(-0.008, ~0.136)$\\
\addlinespace
PWEPH & Vague & 0.525 & 0.033 & 0.472 & 0.034 & 0.053 & 0.047 & 0.868 & $(-0.042, ~0.145)$\\
PWEPH & BHM & 0.525 & 0.033 & 0.470 & 0.033 & 0.054 & 0.047 & 0.879 & $(-0.038, ~0.146)$\\
PWEPH & CP & 0.525 & 0.033 & 0.471 & 0.033 & 0.054 & 0.047 & 0.876 & $(-0.038, ~0.146)$\\
PWEPH & LEAP & 0.524 & 0.029 & 0.457 & 0.030 & 0.068 & 0.042 & 0.942 & $(-0.019, ~0.147)$\\
PWEPH & NPP & 0.535 & 0.027 & 0.455 & 0.028 & 0.080 & 0.039 & 0.979 & $(~0.003, ~0.157)$\\
PWEPH & PP~$(a_0 = 0.5)$ & 0.533 & 0.029 & 0.458 & 0.029 & 0.075 & 0.041 & 0.965 & $(-0.006, ~0.155)$\\
PWEPH & PSIPP & 0.515 & 0.026 & 0.431 & 0.026 & 0.084 & 0.037 & 0.989 & $(~0.012, ~0.155)$\\
\bottomrule
\end{tabular}
}
\vspace{2mm}

\caption*{The columns under IFN and Control show the posterior mean and standard deviation (SD) of the two-year RFS probabilities for the IFN and control arms, respectively. For a given combination of prior and model type (PWEPH or CurePWEPH), the posterior summaries for each treatment arm are based on the model-prior pair with the highest elpd values among models with $J = 2$ to 9 intervals. The corresponding $\Delta$ is computed as the difference in two-year RFS probabilities between the IFN and Control arms using the selected model-prior pairs.}
\label{tab:post_summ}
\end{table}

To quantify the amount of external information incorporated, we compute the effective sample size (ESS) for each model-prior combination, following the definition of ESS in \citet{pennelloExperienceReviewingBayesian2007}. For a given outcome model $M$ and prior distribution $\pi$, the ESS is defined as
$$\text{ESS} = n \cdot \frac{V_{(M, ~\pi_{v})}}{V_{(M, ~\pi)}},$$
where $n$ is the sample size of the current trial, $V_{(M, ~\pi_v)}$ denotes the posterior variance of the treatment effect under model $M$ with the vague prior $\pi_v$, and $V_{(M, ~\pi)}$ denotes the posterior variance under the same outcome model $M$ with prior $\pi$ (which may be informative). Intuitively, the ESS represents the number of subjects that would be required in the current trial, without incorporating external information, to achieve the same posterior variance as obtained when leveraging the external study. Hence, the difference $(\text{ESS} - n)$ can be interpreted as the effective number of subjects borrowed from the external study. 

\begin{figure}
    \centering
    \includegraphics[width=0.9\linewidth]{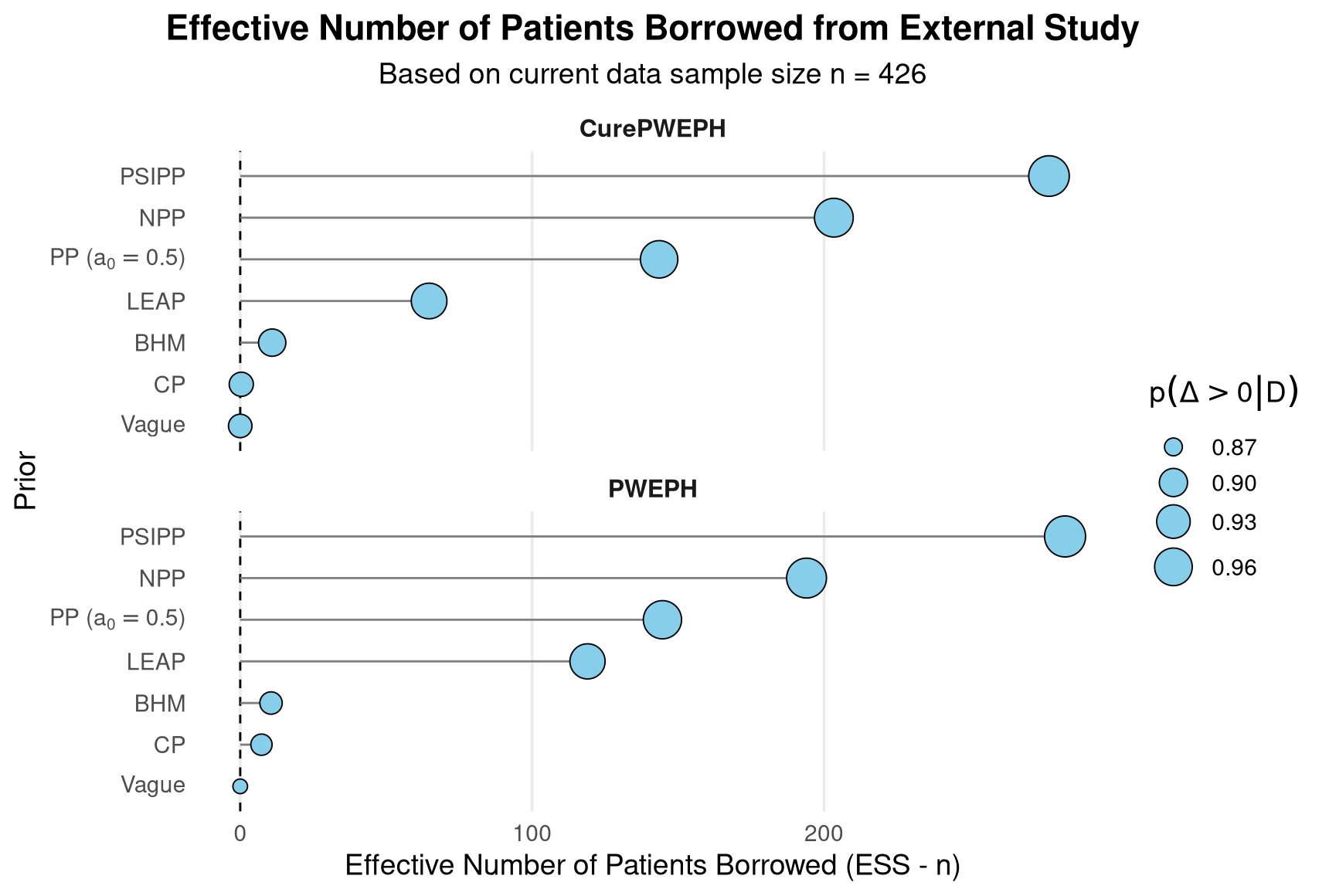}
    \caption{\textbf{Effective number of patients borrowed from the external study, defined as $\text{ESS} - n$ with $n = 426$ for the current trial, across different priors under the CurePWEPH and PWEPH models}. For each model type and treatment arm, the number of intervals $J$ was fixed to the value that maximized the elpd under the vague prior. The size of each point reflects the posterior probability $p(\Delta > 0 \mid D)$.}
    \label{fig:ess_plot}
\end{figure}

To make a fair comparison of the amount of external information borrowed, we fix the outcome model by choosing the number of intervals $J$ that maximizes elpd under the vague prior for each model type within each treatment arm. This ensures that differences in $(\text{ESS} - n)$ reflect the choice of prior rather than variation in the underlying outcome model.
Specifically, we set $J = 4$ for the IFN arm and $J = 5$ for the control arm under the CurePWEPH model, and $J = 5$ for both treatment arms under the PWEPH model.
Figure \ref{fig:ess_plot} presents the effective number of subjects borrowed from the external study for each prior under these selected models.
For both the CurePWEPH and PWEPH models, PSIPP, NPP, PP, and LEAP borrow substantially more information from the external study than the other priors. The BHM and CP yield $(\text{ESS} - n)$ values that are nearly identical to those under the vague prior, indicating minimal borrowing.
This is partly a consequence of how these two priors are specified here (Tables~\ref{tab:supp_PWE_summary} and~\ref{tab:supp_CurePWE_summary}): only the regression coefficients $\bm{\beta}$ are shared between the current and external models, while the baseline hazards and, under the CurePWEPH model, the cure probabilities have independent priors. Since the models are fitted separately by treatment arm, the arm-specific RFS probabilities, and hence $\Delta$, are determined largely by the baseline hazards and cure probabilities, which are not borrowed under the BHM and CP.
The PSIPP, NPP, PP, and LEAP also produce higher posterior probabilities of a positive treatment effect than the other priors. These findings are consistent with the results reported in Table \ref{tab:post_summ}, and the amount of borrowing for each prior remains relatively stable across different choices of the number of intervals (see Supplementary Material).

\begin{figure}
    \centering
    \includegraphics[width=0.8\linewidth]{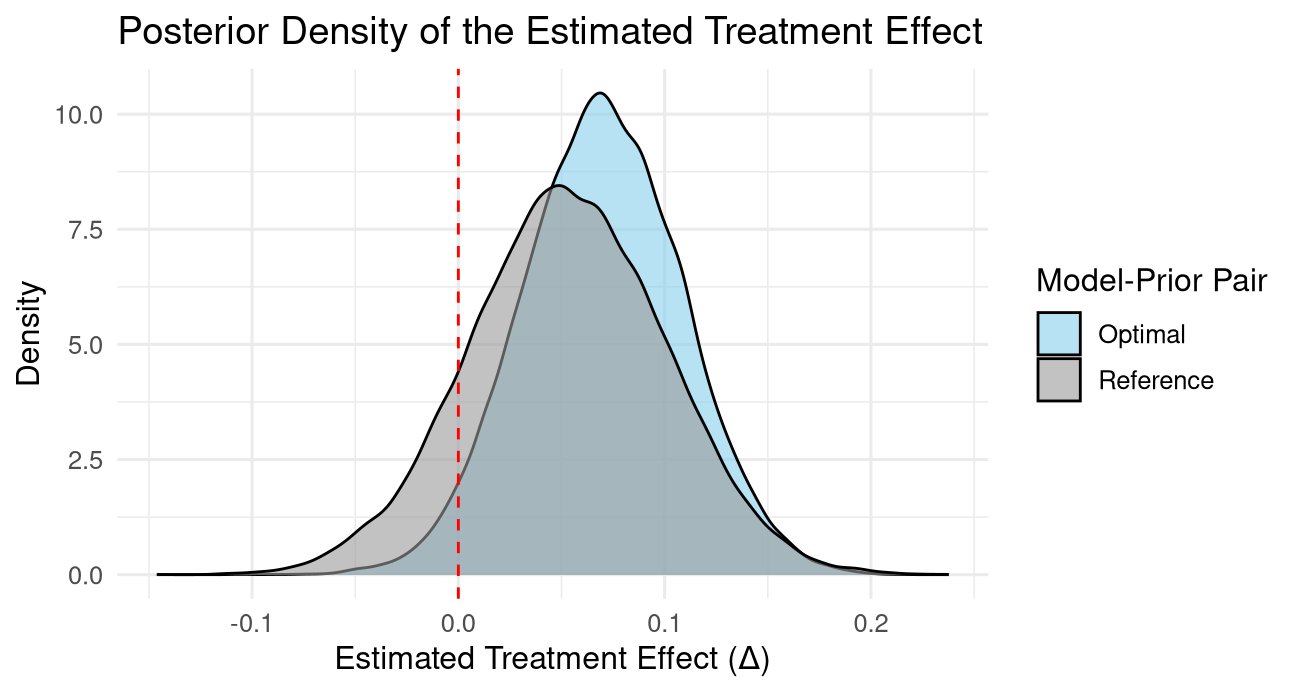}
    \caption{\textbf{Posterior densities of the estimated treatment effect $\Delta$ under the optimal and reference model-prior pairs}. The optimal pair corresponds to the model-prior combination that attains the highest elpd within each treatment arm, while the reference pair is defined analogously but restricted to the vague prior. The vertical dashed line at $\Delta = 0$ indicates the null value of no treatment effect.}
    \label{fig:post_dens_diff}
\end{figure}

Beyond selecting $J$ within each combination of prior and model type, elpd can also be used across combinations to identify the best-performing one within each treatment arm, which we refer to as the optimal model-prior pair: the CurePWEPH model with $J = 2$ intervals under the PSIPP for the IFN arm and the PWEPH model with $J = 5$ intervals under the NPP for the control arm. This specification yields an estimated treatment effect of $\Delta \approx 0.070$ with posterior standard deviation 0.038, a $95\%$ BCI of $(-0.005, 0.146)$, and posterior probability $p(\Delta > 0 \mid D) \approx 0.967$. For comparison, we define a reference model-prior pair analogously by restricting to the vague prior, which corresponds to the PWEPH model with $J = 5$ intervals for both treatment arms. Figure \ref{fig:post_dens_diff} shows the posterior densities of $\Delta$ under the optimal and reference model-prior pairs. Compared to the reference, the posterior distribution under the optimal pair is more concentrated and shifted to the right, indicating stronger evidence of a positive treatment effect when external information is incorporated, whereas the distribution under the reference pair places more mass near the null value of no effect.
Because both the outcome model and the prior in the optimal pair are selected using the current data, separately in each arm, this analysis may be regarded as exploratory; in a confirmatory analysis they would be pre-specified~\citep{FDA2023covariates}.

%% file: 003_2-logistic.tex
\subsection{Bayesian logistic regression with informative priors}
\label{sec:log_reg}

We now showcase how to employ external data and dynamic borrowing priors to perform important statistical tasks such as estimating causal effects and doing variable selection in a binary regression context.
The analysis of binary data provides a contrast with the survival analysis in the previous section and helps highlight the broad applicability of DBPs and external data methods in general.

Throughout this section, our main goal will be to analyze data on two clinical trials on progression to AIDS in a cohort of HIV-infected patients treated with zidovudine (AZT) using logistic regression as in \citet{Chen1999b} (Section 4.2 therein).
The historical data are from ACTG019~\citep{Volberding1990}, a double-blind trial of AZT against placebo in asymptomatic HIV-infected persons with fewer than $500$ CD4 cells/mm$^3$; $n_0 = 822$ patients have complete data.
The current data are from ACTG036~\citep{Merigan1991}, a placebo-controlled trial of AZT in asymptomatic HIV-infected patients with hemophilia, with $n = 183$ observations.
The outcome equals 1 if the patient died or developed AIDS or AIDS-related complex (ARC), and 0 otherwise.
We will use the following covariates: CD4 cell count (\texttt{cd4}), \texttt{age}, \texttt{treatment} and \texttt{race}.
The continuous covariates (\texttt{age} and \texttt{cd4}) are centered and scaled, which keeps the regression coefficients on comparable scales and facilitates computation; we recommend this step in general.
The interested reader is referred to Section 4.2 of \citet{Chen1999b}. 

Letting $y_i = 1$ if the $i$-th individual presented the (composite) negative outcome and $y_i = 0$ otherwise and letting $\boldsymbol{X}_i$ stand for the set of covariates for this individual, our base model is the usual Bayesian logistic regression:
\begin{align*}
   \operatorname{logit}\left(\operatorname{Pr}\left(y_i = 1 \mid \boldsymbol{X}_i, \boldsymbol{\beta}\right)\right)  &= \boldsymbol{X}_i\boldsymbol{\beta}
   , \qquad 
    \boldsymbol{\beta} \sim \operatorname{MVN}\left(\boldsymbol{0}, \operatorname{diag}\left(\boldsymbol{\sigma}_b\right)\right),
\end{align*}
where we initially pick a multivariate normal prior as our initial prior $\pi_0$ for the coefficients with a diagonal covariance matrix.
This particular choice allows us to pick different scales for the intercept and the covariate effects, which can be helpful in getting the scales right.
Another alternative is the so-called weakly informative (Cauchy) priors of \citet{Gelman2008}, which we discuss below when considering prior sensitivity analyses -- see Section~\ref{sec:PSA}.

The incorporation of historical data is likely to increase precision in the analysis because $\pi_0$ will be updated with the information from the historical data $D_0 = (\boldsymbol{y}_0, \boldsymbol{X}_0)$ and thus help us better estimate the effect conditional on the current data $D = (\boldsymbol{y}, \boldsymbol{X})$.
Treatment effects are not always exchangeable across studies \emph{marginally}, i.e.\ without conditioning on covariates.
Our formulation thus makes a \emph{conditional} exchangeability assumption and employing logistic regression allows for controlling for prognostic covariates and thus improving efficiency in estimating the \textbf{treatment effect}, which is the main concern.
However, including too many covariates can impact efficiency because if a covariate is not relevant to explain the outcome it might add noise.

\subsection{Variable selection}
\label{sec:varsel}

A natural question therefore arises: which predictors should be included?
At one extreme, one could attempt to include all possible predictors.
In some situations, this may not be possible due to intractability or correlations between covariates.
For binary outcomes, including too many predictors can result in complete separation -- where a predictor or a combination of predictors perfectly predicts the outcome.
Even if it is possible, including predictors that are not relevant can inflate posterior variances, resulting in an inefficient estimate.
At the other extreme, one could fit all possible models and choose the one that results in the best model fit according to some model selection criterion (e.g., AIC or BIC).
This approach, however, ignores model uncertainty and may tend to underestimate the variance in the effect estimate(s).

The FDA guidance on covariate adjustment in randomized trials \citep{FDA2023covariates} states that, while an unadjusted analysis remains acceptable for the primary endpoint, adjustment for prognostic baseline covariates is generally acceptable.
This will typically reduce the variance of the estimated treatment effect, provided the covariates and the full analysis procedure are prespecified before any unblinding of comparative data.
For nonlinear models with non-collapsible estimands, the guidance requires that the estimand be declared as conditional or unconditional and recommends that unconditional effects be estimated by methods such as g-computation or inverse probability of treatment weighting, with standard errors that account for stratified randomization.

Our goal is to identify the best subset of models that captures the relationship between the response variable and the predictors in a simple yet effective way, as discussed in \citet{Mitchell1988}.
Several methods have been proposed to address this problem, but computational challenges arise when the number of predictors is large, as noted by \citet{George1993}. In this context, Bayesian subset selection offers a powerful framework for handling high-dimensional settings.

Once more, we will analyze the AZT treatment example in \citet{Chen1999b} (Section 4.2 therein).
In order to perform model selection, we will compute marginal likelihoods and employ Bayesian model averaging (BMA) to obtain marginal estimates of relevant quantities over the space of possible models.
Let $k$ denote the maximal number of covariates (i.e. the full model) and let $\mathcal{M}$ be the set of the $2^k - 1$ models that include at least one covariate (the intercept-only model is not considered).

\subsubsection{Distributions on the model space}
\label{sec:dist_model_space}

BMA requires specification of (1) \emph{prior} model probabilities $\{ p(m), m = 1, \ldots, M \}$ reflecting our uncertainty about the true model, (2) \emph{likelihoods} $\{ L_m, m = 1, \ldots, M \}$ for each model, and (3) \emph{priors} $\{ \pi_m, m = 1, \ldots, M \}$ for parameters for each likelihood.
Armed with these, BMA proceeds by computing \emph{posterior} model probabilities \textit{via}
\begin{align}
    \label{eq:model_post}
    p(m|D) \equiv q(m|D_0, D) =  \frac{ Z_m(D) p(m) }{ \sum_{j \in \mathcal{M}} Z_{j}(D) p(j) }, \ \ m = 1, \ldots, |\mathcal{M}|
\end{align}
where $Z_m(D)$ is the marginal likelihood for model $m$ and we drop the dependence on $D_0$ for notational convenience.
These posterior probabilities quantify the plausibility of each model based on the prior and the data.

Perhaps the most difficult part of the prior elicitation in BMA is the prior model probabilities.
However, given historical data, this is made easy. Specifically, we compute the prior model probabilities using the marginal likelihoods of the external data, i.e.,
\begin{align*}
p(m) &\equiv q(m \mid D_0) = \frac{\widetilde{Z}_m(D_0)}{ \sum_{j \in \mathcal{M}} \widetilde{Z}_j(D_0) },
\end{align*}
where $\widetilde{Z}_m(D_0) = \int L(\bm{\beta}^{(m)} | D_0) \pi_0 \left(\bm{\beta}^{(m)} \mid \phi_0 \right) d\beta^{(m)}$ is the marginal likelihood using external data set $D_0$ and prior $\pi_0$, which is specified to be $\bm{\beta}^{(m)} \sim N(\bm{0}, \phi_0^2 \cdot \bm{I}_{p^{(m)}})$, where $\phi_0 = 1$ and $p^{(m)}$ is the dimension of $\bm{\beta}^{(m)}$. Notice that Equation \ref{eq:model_post} simplifies to $\frac{ Z_m(D) \widetilde{Z}_m(D_0) }{ \sum_{j \in \mathcal{M}} Z_{j}(D) \widetilde{Z}_j(D_0) }$. 

For each model $m \in \mathcal{M}$, we elicit a normalized power prior for the regression coefficients $\bm{\beta}^{(m)}$.
The initial prior is the same density as that described previously.
We further specify $a_0 \sim \text{Beta}(\delta_0 = 1, \lambda_0 = 1)$.
This yields the joint prior
\begin{align*}
\pi\left(\beta^{(m)}, a_0 \mid D_0\right) &\propto \frac{L\left(\beta^{(m)} \mid D_0\right)^{a_0}}{C(a_0)} a_0^{\delta_0-1}(1-a_0)^{\lambda_0 - 1} \pi_0\left(\beta^{(m)} \mid \phi_0\right), 
\end{align*}
where $C(a_0) = \int L(\beta^{(m)} | D_0)^{a_0} \pi_0\left(\beta^{(m)} \mid \phi_0\right) d\beta^{(m)}$.

Table \ref{tab:ml_var_sel} shows the results for this data analysis. The first three rows comprise approximately 90\% of the posterior mass, indicating that our inferences are going to be mostly influenced by these models, which are, (i) age, treatment, cd4, (ii) age, treatment, race, cd4, and (iii) treatment, cd4. Treatment and cd4 are included in all three.
However, the posterior model probabilities are close to the prior model probabilities $p(m \mid D_0)$, which use ACTG019 at full weight.
Among the models that include cd4, the log marginal likelihoods $\log Z_m(D)$ differ by less than $0.2$, and the largest value is attained by M7 (cd4 only).
The marginal likelihood of the current data therefore does not favor the models that include treatment, and their high posterior probability derives mainly from the historical data through $p(m \mid D_0)$.

\begin{table}[htbp]
\caption{\textbf{Model probabilities for the AZT treatment example.} The table displays the set of considered submodels along with their corresponding posterior probabilities. Models are ordered by decreasing posterior probability.}
\centering
\begin{tabular}{lcccc}
  \hline
    $m$ & Covariates & $p(m \mid D_0)$ & $\log(Z_m(D))$ & $p(m | D)$ \\ 
  \hline
  M1 & age, treatment, cd4 & 0.540 & -35.70 & 0.557 \\ 
  M2 & age, treatment, race, cd4 & 0.254 & -35.79 & 0.238 \\ 
  M3 & treatment, cd4 & 0.100 & -35.70 & 0.103 \\ 
  M4 & treatment, race, cd4 & 0.049 & -35.81 & 0.045 \\ 
  M5 & age, cd4 & 0.032 & -35.68 & 0.033 \\ 
  M6 & age, race, cd4 & 0.015 & -35.77 & 0.014 \\ 
  M7 & cd4 & 0.006 & -35.65 & 0.007 \\ 
  M8 & race, cd4 & 0.003 & -35.76 & 0.003 \\ 
  M9 & age, treatment & 0.000 & -41.46 & 0.000 \\ 
  M10 & age, treatment, race & 0.000 & -41.58 & 0.000 \\ 
  M11 & treatment & 0.000 & -41.84 & 0.000 \\ 
  M12 & treatment, race & 0.000 & -41.96 & 0.000 \\ 
  M13 & age & 0.000 & -41.67 & 0.000 \\ 
  M14 & age, race & 0.000 & -41.79 & 0.000 \\ 
  M15 & race & 0.000 & -42.11 & 0.000 \\ 
   \hline
\end{tabular}
\label{tab:ml_var_sel}
\end{table}

\subsubsection{Estimating the average treatment effect}
\label{sec:ate}

The analysis thus far has yet to answer what is the treatment effect, i.e., what is the impact of the treatment, AZT, on the probability of death/negative progression compared to placebo?
In the sequel, we use Bayesian $g$-computation techniques~\citep{Keil2018} to compute this marginal effect.
To evaluate the causal impact of treatment in the AZT example, we begin by analyzing the ATE with a frequentist approach, using empirical means and generalized linear models to compare the results.
First, we compute the empirical ATE (the difference between the average response in the treatment and control groups): $-0.0296$.
Then, we fit a logistic regression model in two scenarios: with the treatment as the only covariate and with all covariates.
In order to estimate the ATE within the Bayesian framework, we begin by defining the model space and establishing prior distributions for model parameters as in Section~\ref{sec:dist_model_space}, incorporating external data through likelihood functions and prior beliefs.
Then, we use a weighted combination of the posterior probabilities presented in Table~\ref{tab:ml_var_sel} with Dirichlet-weighted averaging to compute the marginal means providing a comprehensive view of the treatment effect following \citet{rubinBayesianBootstrap1981}.
The results are presented in the Supplementary Material.

We take a Bayesian bootstrap approach~\citep{rubinBayesianBootstrap1981,Wang2015} -- see for instance, Section 3.3 of~\citet{Oganisian2021}.
The algorithm proceeds as follows:

\textbf{Algorithm to compute risk difference for each model.}
Let $\bm{\beta}^{(m)} = \left(\beta_0^{(m)}, \beta_1^{(m)}, \bm{\beta}_2^{(m)} \right)$, where, for model $m$, $\beta_0^{(m)}$ is the intercept, $\beta_1^{(m)}$ is the conditional treatment effect (which is equal to zero when treatment is excluded from a model), and $\bm{\beta}_2^{(m)}$ contains the regression coefficients for the non-treatment covariates.
\begin{enumerate}
    \item Sample from the posterior density for model $m$, i.e., $p(\bm{\beta}^{(m)} | D, D_0, M = m)$;
    \item For each treatment arm $a = 0, 1$, compute
    $$
        \mu_a^{(m)} \equiv \sum_{i=1}^n \omega_i \cdot \text{logit}^{-1}\left( \beta_0^{(m)} + \beta_1^{(m)}a + \bm{x}_i'\bm{\beta}_2^{(m)} \right),
    $$
    where $(\omega_1, \ldots, \omega_n) \sim \text{Dirichlet}(1, \ldots, 1)$;
    \item Compute the risk difference $\Delta^{(m)} = \mu_1^{(m)} - \mu_0^{(m)}$. This returns a sample from $p(\Delta^{(m)} | D, D_0)$.
\end{enumerate}
Repeating steps (1)--(3) for many MCMC samples across all models results in a posterior distribution for the risk difference for each model. 

\begin{figure}[htbp]
    \centering
    \includegraphics[scale=.4]{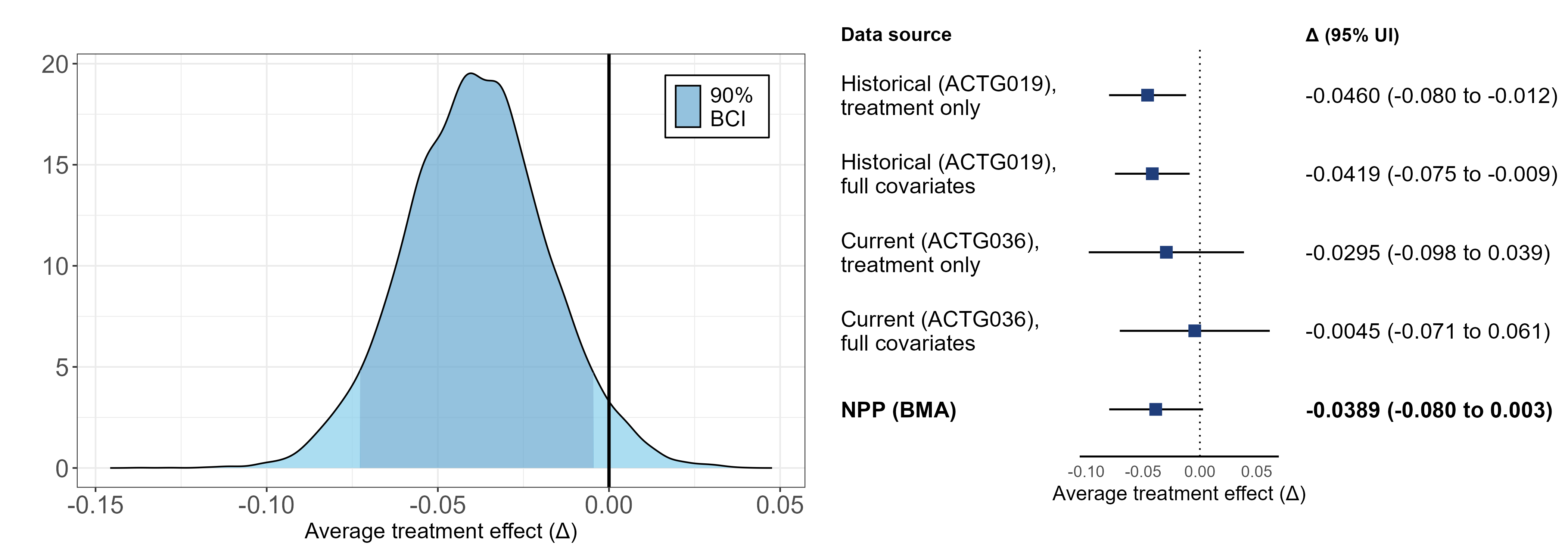}
    \caption{\textbf{Posterior distribution of the ATE}. On the \textbf{left} panel we show the posterior density for the average treatment effect (ATE), given by the difference $\mu_1 - \mu_0$ of the model-averaged expected probabilities for each trial arm -- control and treatment -- the solid vertical line marks ATE = 0.
    The \textbf{right} panel displays the ATE estimated using logistic regression via maximum likelihood under each setting with its $95\%$ confidence interval, compared to the ATE estimated using BMA.
    }
    \label{fig:ate_posterior}
\end{figure}

Finally, the model-averaged posterior risk difference may be computed as a weighted average of the model-specific posteriors, i.e.,
$$
    p(\Delta \mid D, D_0) \;=\; \sum_{m \in \mathcal{M}}
  p(m \mid D, D_0)\, p(\Delta^{(m)} \mid D, D_0).
$$
To obtain posterior samples from the model-averaged distribution, one can (1) sample $z$ from a categorical distribution with probabilities according to the posterior model probabilities and (2) sample from the posterior distribution of $\Delta^{(z)}$, repeating this process many times.
Figure \ref{fig:ate_posterior} displays the samples obtained by this procedure.
Notice how the 90\% posterior BCI for the ATE, averaged over all models, excludes zero (left panel), whereas the 95\% BCI, $(-0.080, 0.003)$, includes it (right panel and Table~\ref{tab:ate_glm}).
In both cases the posterior still conveys that there is substantial uncertainty about the actual value of the ATE and odds ratio.


\subsubsection{Prior sensitivity analysis}
\label{sec:PSA}


While we have so far argued for the use of Bayesian external data methods by showing the gains in efficiency, a major issue in applied Bayesian analysis is assessing to which extent the results depend on the chosen prior~\citep{Lopes2011,Kallioinen2024}.
When employing external data in order to create an informative prior, the issue of sensitivity to the prior gains even more preeminence.
We thus show how one can conduct a prior sensitivity analysis (PSA) in the context of DBPs.
We start by conducting a sensitivity analysis for the normal prior by varying its hyperparameters.
For simplicity, all of the analysis in this section will be performed under the normalized power prior (NPP), unless otherwise stated.

First, notice that we can rewrite the model posterior probability as
\begin{align*}
    p(m \mid D) 
    &= \frac{\int L(\beta^{(m)} | D) \pi\left(\beta^{(m)}, a_0 \mid D_0\right) d\beta^{(m)} da_0 \int L({\beta}^{(m)} | D_0) \pi_0 \left({\beta}^{(m)} \mid \phi_0 \right) d\beta^{(m)} }{\sum_{j \in \mathcal{M}} \int L({\beta}^{(j)} | D) \pi\left(\beta^{(j)}, a_0 \mid D_0\right) d\beta^{(j)} da_0 \int L({\beta}^{(j)} | D_0) \pi_0 \left({\beta}^{(j)} \mid \phi_0 \right) d\beta^{(j)} }.
\end{align*}

\begin{figure}
    \centering
    \includegraphics[scale=.4]{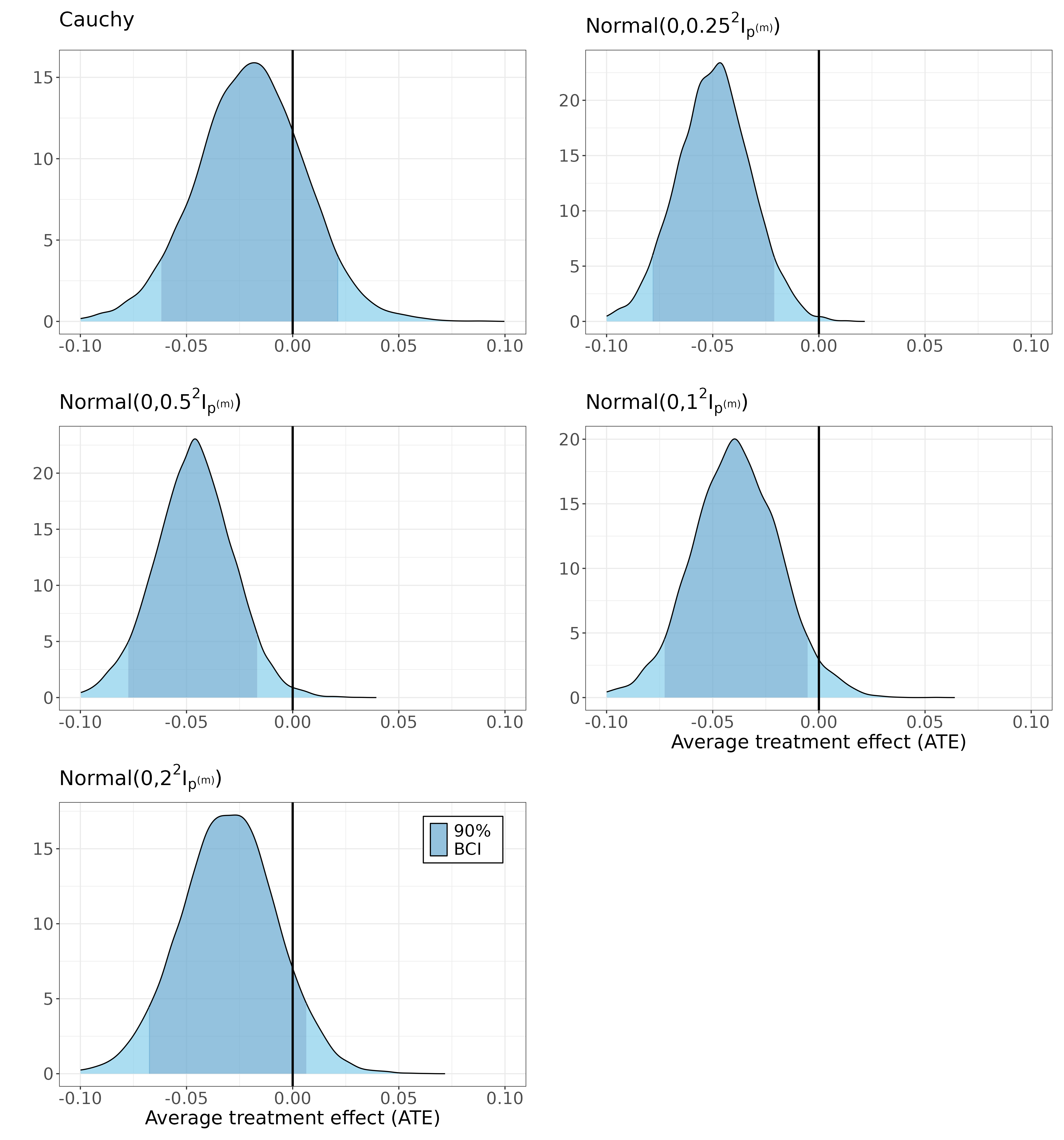}
    \caption{\textbf{Posterior distributions of the average treatment effect (ATE) for each choice of initial prior.} We show the posterior density along with 90\% Bayesian credibility intervals (BCI) for the ATE under normal priors with different scales as well as weakly-informative Cauchy priors as the initial prior ($\pi_0(\boldsymbol{\beta} \mid \phi_0)$, see text).}
    \label{fig:ate_priors}
\end{figure}

From this,  we can see that $Z_m(D)$ and $Z_m(D_0)$ depend on the parameter $\phi_0$ and that the former depends on the parameters $\delta_0$ and $\lambda_0$.
As described in Section 4.2 of \citet{Chen1999b}, we performed a sensitivity analysis for $\phi_0$ by computing model probabilities for several choices of parameters.
We fixed $\delta_0 = \lambda_0 = 1$, and computed the probabilities for several values of $\phi_0$ as shown in the left column of Figure~\ref{fig:model_prob_before_after_PSM}.
Our choices for the hyperparameter ($\phi_0$) reflect progressively `flatter' priors for the coefficients, but we take care not to make $\phi_0$ too large, because the linear predictor in this model is in logit space.
The interested reader is referred to \citet{Seaman2012} and references therein.
We can see that the probabilities become less concentrated as $\phi_0$ increases and the model with the highest probability changes to the full model for small values of $\phi_0$.
This sensitivity of posterior model probabilities (and Bayes factors) to prior specification is well-known -- see Section 5 of \citet{Kass1995} and also \citet{Sinharay2002} -- and is an almost inescapable part of Bayesian model selection.
The goal of our analysis is to show such sensitivity can be investigated in an external data scenario.
In addition, Figure \ref{fig:ate_priors} shows that the $90\%$ credible interval starts to include $0$ as $\phi_0$ increases, which is in line with intuition: further uncertainty in the prior gets carried over to the posterior distribution. 

\cite{Gelman2008} discusses weakly informative priors (WIP) for logistic regression.
As a point of comparison, we performed a prior comparison by using a Cauchy weakly informative initial prior.
We set a Cauchy$(0, 10)$ prior for the intercept and Cauchy$(0, 2.5)$ for the other coefficients.
We can see in the left column of Figure~\ref{fig:model_prob_before_after_PSM} that the Cauchy prior assigns higher prior and posterior probabilities to the full model compared to the Normal prior.
Additionally, it assigns more distributed probabilities: for example, M1 has twice the prior probability in the Normal case, while M3, M4 and M5 have similar prior and posterior probabilities for the Cauchy prior.
Also, the Cauchy prior results in a posterior distribution for the ATE which is more concentrated around $0$, as shown in Figure \ref{fig:ate_priors}. 


Besides model probabilities, one might wonder how different  prior distributions affect other quantities, such as the predictive probabilities, i.e., one might be interested in checking what kind of data the prior distribution implies -- see \citet{Gelman2017} for a thorough discussion.
As a practical Bayesian visualization discussed in \citet{Gabry2000}, we performed prior predictive checking (PPC) to compare the initial priors (Normal with various scales and Cauchy).
We also employed various beta priors on the discounting parameter, $a_0$.
We started by computing the prior predictive distributions of the outcome probability under these different priors under model M1 (age, treatment, cd4), i.e., the model with the highest probability in Table \ref{tab:ml_var_sel}.
Figure \ref{fig:ppc_all} provided in the Supplementary Material presents the results for this analysis, which show that the WIP can lead to substantially different predicted probabilities.
In contrast, results seem less sensitive to the prior on the discounting parameters, as shown in the bottom panels of Figure \ref{fig:ppc_all} of the Supplementary Material.

For binary data of the type considered in this section, receiver operating characteristic (ROC) curves are especially suited, because they show the predictive power of the model under consideration in an interpretable way.
We computed \textit{prior receiver operating characteristic} (pROC) curves  and associated area under the curve (AUC) by using the \textbf{current} data as the gold standard and looking at the prior predictive probabilities as informed by the historical data.
The results of these analyses are shown in the Supplementary Material in Figures~\ref{fig:roc_beta} and~\ref{fig:roc_a0} and indicate that in terms of predicting the actual binary outcome, the priors usually lead to pretty similar performance, at least \textit{a priori}.
A notable exception is the WIP Cauchy  prior, which leads to a higher AUC.

%% file: 003_3-method_comp.tex
\subsection{Comparing different dynamic borrowing priors}
\label{sec:method_comparison}

Finally, we compare the dynamic borrowing priors (DBPs) described in Section~\ref{sec:multimodels}, along with some baseline priors we shall discuss shortly with regard to the posterior distribution of the odds ratios.
For comparison we also perform analyses where we (i) pool the historical and current data together and perform a standard logistic regression with standard normal priors on the coefficients and (ii) disregard the historical data and analyze only the current data under different priors.
For (ii) we used (a) a standard normal prior on $\boldsymbol{\beta}$, which we will call a `containment' prior, designed to keep effects close to 0 on the logit scale; (b) a flat (independent) normal prior with standard deviation $10$, which we will call `flat'; and (c) a weakly informative prior (WIP) using the Cauchy as previously described.
Figure~2 of \citet{Alt2025} compares the posterior estimates of the regression coefficients under the borrowing priors for these data, on the logit scale.
Here we report the estimates on the odds ratio scale and add analyses of the current data alone under three priors (containment, flat and weakly informative), and of the pooled data, so that the borrowing priors can be compared with each other and with the effect of the prior choice when no external data are used.

\begin{figure}[!ht]
    \centering
    \includegraphics[scale=.4]{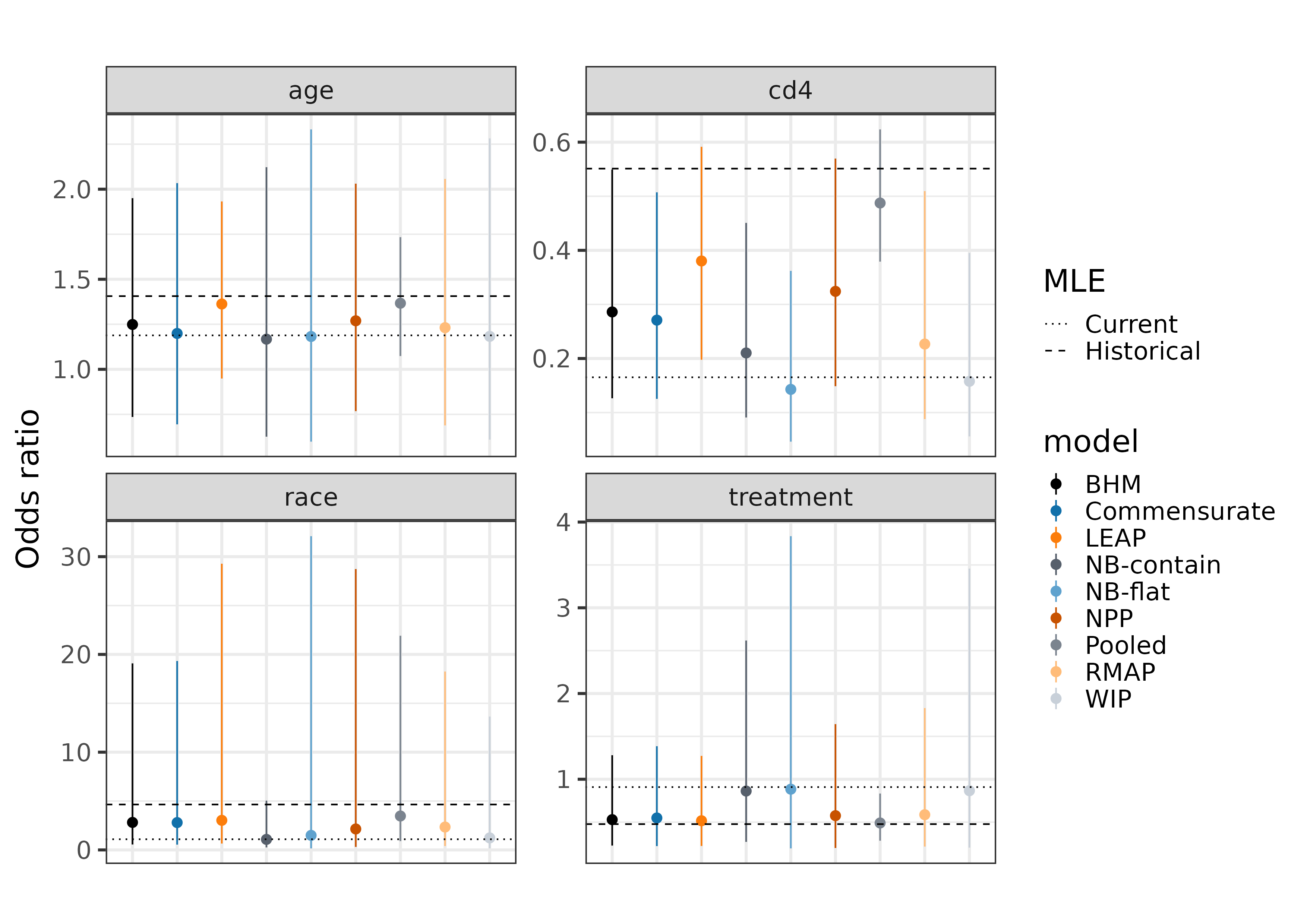}
    \caption{\textbf{Posterior mean and 95\% credibility interval for the conditional odds ratio of each covariate under different external data models}.
    For each covariate $j$ we show the posterior mean and 95$\%$ BCI odds ratio $\operatorname{OR}_j = \exp(\beta_j)$ obtained under each of the five external data borrowing methods (NPP, BHM, rMAP, Commensurate and LEAP) considered and also various choices of priors when disregarding the historical data (see text).
    In addition, we show the Bayesian posteriors obtained by either analyzing just the current data (no borrowing) and also by combining current and external data as one (pooling).
    In the legend, NB-contain, NB-flat and WIP denote analyses of the current data only (no borrowing) under the $N(0, 1)$ (``containment''), $N(0, 10^2)$ (``flat'') and Cauchy weakly informative priors, respectively; Pooled denotes the analysis of the combined current and historical data with $N(0, 1)$ priors; RMAP denotes the rMAP prior. The borrowing priors use the specifications of \citet[Section~4.1]{Alt2025}, whose Figure~2 compares them for the same data on the logit scale; that figure does not include the analyses without borrowing or the pooled analysis.
    The maximum likelihood estimates for each data set are shown as horizontal dashed/dotted lines (dotted: ACTG036, current data; dashed: ACTG019, historical data).}
    \label{fig:method_comparison}
\end{figure}

The results displayed in Figure~\ref{fig:method_comparison} show that employing external data through DBPs does indeed increase estimation efficiency, as evidenced by the narrower BCIs obtained under the Commensurate, BHM, LEAP, rMAP and NPP priors.
Within the DBPs we also observe different posterior distributions, notably the rMAP leading to wider intervals when compared to, e.g. the NPP and more extremely to the LEAP.
This indicates different DBPs have different borrowing properties, which deserves thorough future investigation.

Unsurprisingly given the differences in sample size ($n_0=822$ vs $n=183$) pooling the two datasets leads to a rather narrow BCI centered around the maximum likelihood estimate for the historical data, which is much larger in this example.
On the other hand, analyzing the current data only also leads to different posterior distributions depending on the employed prior.
For instance, the WIP (light blue in Figure~\ref{fig:method_comparison}) leads to rather large uncertainty on the posterior for the treatment OR.
These results once again reinforce that this analysis is a situation in which one would like to give the prior special attention.

%% file: 003_4-PSM.tex
\subsection{Propensity-score matching: further restricting external data influence}
\label{sec:propensity}

As discussed in Section \ref{sec:ps}, propensity‐score–based approaches can be used to identify subsets of external data that are most compatible with the current study population before incorporating them into the Bayesian analysis. Building on this framework, we examined how the analysis results changed after performing propensity score matching (PSM) on the external data.
In the logistic regression example, the external data set contained $n_0 = 822$ subjects, which was much larger than the current study sample ($n = 183$).
Because the external and current populations may differ in their baseline characteristics, it is reasonable to borrow information only from external subjects who are most comparable to those in the current study.
PSM provides a straightforward way to identify such a subset and serves as a useful sensitivity analysis.
The matching-assisted power prior of \citet{Qian2025} follows the same two steps, selecting external controls by matching and then borrowing from them through a power prior; here we combine matching with the NPP and examine how the results change.

Specifically, we combined the current and external data sets and defined a binary indicator $S_i$, where $S_i = 1$ if subject $i$ belongs to the current study and $S_i = 0$ otherwise. The propensity score (PS), $e_i = \Pr(S_i = 1 \mid \bX_i = \bm{x}_i)$, represents the probability that subject $i$ comes from the current study given their covariates $\bm{x}_i$ (here, \texttt{cd4}, \texttt{age}, and \texttt{race}).
To flexibly model nonlinear effects and interactions among covariates, we estimated the PS using Bayesian additive regression trees (BART, \cite{chipmanBARTBayesianAdditive2010}), a Bayesian nonparametric method that models an unknown function $f(\bm{x})$ using a sum of regression trees:
\begin{align*}
    S \mid \bX = \bm{x} &\sim \text{Bernoulli}\left[\Phi\{f(\bm{x})\}\right], \\
    f(\bm{x}) &= \sum_{j=1}^{m} g(\bm{x}; \mathcal{T}_j, \mathcal{M}_j),
\end{align*}
where $S$ is the binary indicator for belonging to the current study, and $\Phi(\cdot)$ denotes the cumulative distribution function (CDF) of a standard normal random variable. Each $\mathcal{T}_j$ represents a binary regression tree, $\mathcal{M}_j$ denotes the terminal node parameters associated with $\mathcal{T}_j$, and $g(\bm{x}; \mathcal{T}_j, \mathcal{M}_j)$ is a function that maps $\bm{x}$ to the corresponding terminal node parameter. BART imposes regularizing priors on both tree structures and terminal node parameters to limit the effect of each individual tree. This Bayesian framework yields posterior draws of $f(\bm{x})$, and consequently posterior draws of PS. We implemented BART using the R package \textbf{dbarts} \citep{dbarts_package}, taking the posterior mean of the estimated PS for each subject for matching.

Next, we conducted 1:1 nearest neighbor matching between the current and external data based on the logit of these estimated PS using the \textbf{MatchIt} R package \citep{hoMatchItNonparametricPreprocessing2011}.
A caliper of $0.1$ was applied on the logit scale, such that each current subject was matched to the external subject with the closest logit(PS), provided the absolute difference did not exceed $0.1$.
External subjects outside this caliper were excluded from matching.
After propensity score matching (PSM), $103$ external subjects were retained and included in the subsequent analysis. 

\begin{figure}[htpb]
     \centering
     \includegraphics[width=0.93\linewidth]{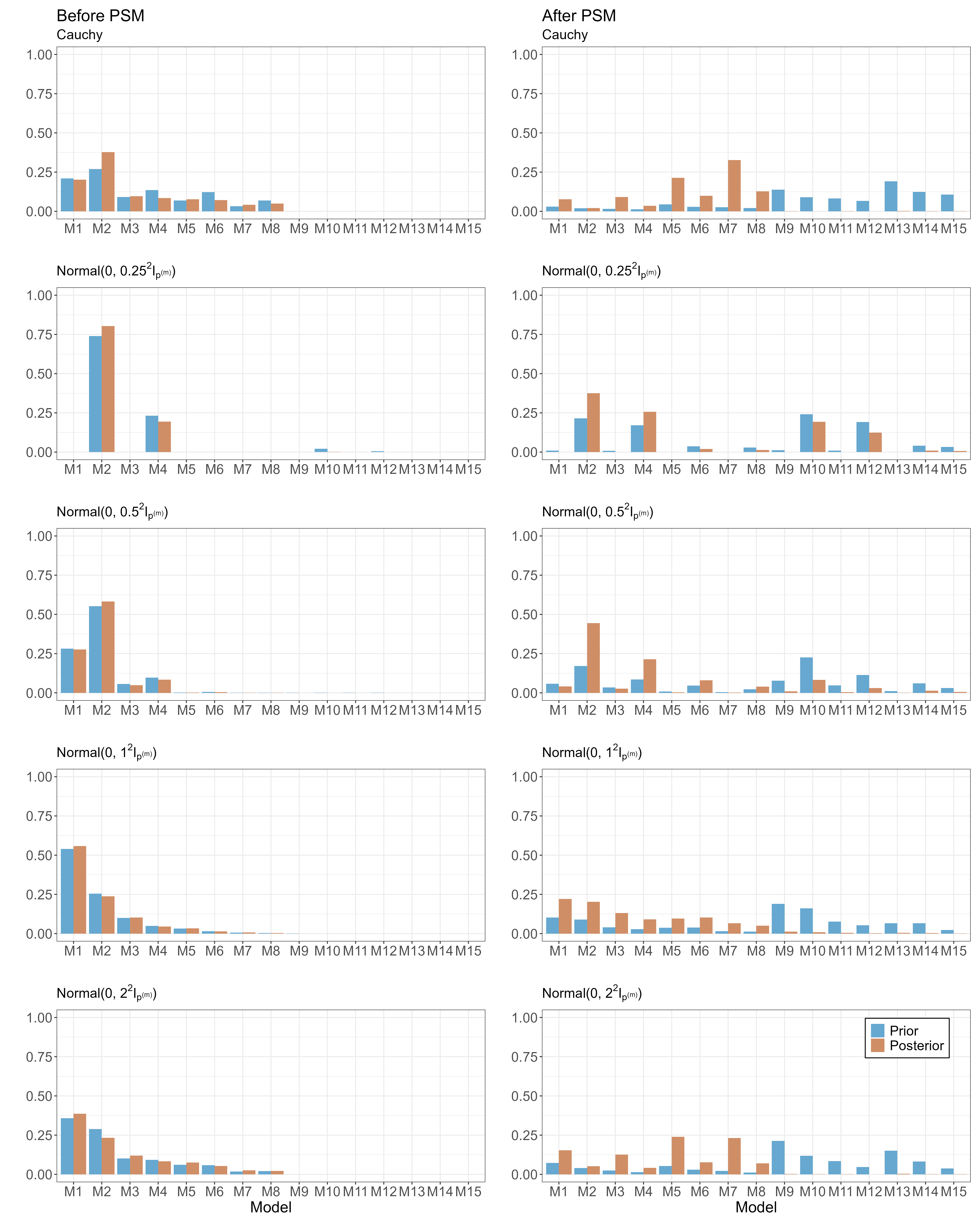}
    \caption{\textbf{Comparison of model probabilities before and after PSM.} Each column displays the model probabilities: on the left we show them before PSM and on the right, after PSM. The Cauchy prior was specified to be Cauchy$(0, 10)$ for the intercept and Cauchy$(0, 2.5)$ for the other coefficients. The models are ordered based on the posterior probabilities of the prior Normal$(0, I_p)$.}
    \label{fig:model_prob_before_after_PSM}
\end{figure}

In Figure~\ref{fig:model_prob_before_after_PSM} we show how model probabilities change before (i.e. without) and after PSM and how they interact with the choice of initial prior for the logistic regression coefficients.
Predictably, the after PSM distributions are substantially more entropic on account of the historical data being substantially reduced ($n^{\texttt{PSM}}_0 = 103$ against the original $n_0 = 822$).
Less intuitively, the models with highest posterior probability also change after PSM, likely reflecting the sensitivity of the estimates to the prior and relatively minor features of the likelihood -- see~\cite{Dellaportas2012}.  

%% file: 004-discussion.tex
\section{Conclusion and future research}
\label{sec:discussion}

Bayesian methods constitute a powerful and principled approach in situations where external (historical) information is available.
In most situations, however,  the analyst would like to borrow information in a controlled manner so as to achieve good inferences without letting the external data dominate the posterior.

In this paper, we have reviewed and discussed practical implementations of the most commonly used historical data borrowing priors (PP, NPP, robust MAP, and commensurate prior) while also discussing recent innovations (e.g., the LEAP and propensity score integrated priors).
We have also discussed how to apply these priors in advanced settings, including survival analysis, covariate adjustment with $g$-computation, and covariate selection.
The main message is that it is possible to incorporate dynamic borrowing techniques into a modern Bayesian statistical workflow and reap the rewards of recent computational and theoretical developments. 

Our study has a few limitations that are worth noting.
First, we do not explore the frequentist characteristics of different DBPs, instead focusing on showcasing how different methods perform when applied on the same real data sets.
The interested reader is referred to \cite{Lee2025,Lee2025b,Lee2026} for more on type I error of Bayesian methods with historical/external data. 
A more thorough theoretical as well as empirical comparison of the borrowing properties of DBPs \citep{Pawel2023} would be an important next step, but is outside the scope of the present study.
We have also not addressed questions of experiment or trial design -- see \cite{Lesaffre2024} and references therein -- or  of operating characteristics (type I and II error, etc.), which are partially addressed in~\cite{Lee2024},~\cite{Lee2025} and~\cite{Lee2025b}.

\subsection{Future research}

Our focus has been on parametric models as they tend to dominate the literature on historical data borrowing.
Nevertheless, Bayesian nonparametric approaches are promising, particularly for regression analysis and density estimation.
However, their use with historical data remains in its infancy.

Another important line of future inquiry is elicitation methods based on only summary data instead of the entire historical dataset.
Due to privacy concerns, working with summary (sufficient) statistics might make it easier for researchers to get access to historical/external data and produce robust estimates at the expense of an extra randomization step \citep{Avella2021}.

Finally, Bayesian analysis with external data faces many of the same questions posed to more traditional Bayesian methods, namely those pertaining to prior sensitivity and robustness to model misspecification.
Despite some efforts on how to specify the hyperpriors for dynamic borrowing (e.g. \cite{Demartino2025} and~\cite{Shen2026}), comparatively little is known on how to specify the prior on the power prior discounting parameter in the NPP or the variance components in the rMAP.
Regarding robustness to model misspecification, the connections to other approaches such as the Safe Bayesian algorithm~\citep{Grunwald2017} or coarsened posterior~\citep{Miller2019} are still underexplored in the literature.

%% file: supp.tex
\section{Computational details for the survival analysis}\label{appendix:surv_comp}

This section provides additional computational details for the survival analysis, including prior specifications for the outcome models and the algorithm used to compute marginal two-year relapse-free survival (RFS) probabilities and the corresponding treatment effect.

Let $D = (\by, \bm{\nu}, \bX, n)$ denote the current data, where $\by = (y_1, \ldots, y_n)'$ is the vector of observed event or censoring times, $\bm{\nu} = (\nu_1, \ldots, \nu_n)'$ is the vector of event indicators with $\nu_i = 1$ for an event and 0 otherwise, and $\bX$ is the $n \times p$ matrix of covariates with $i$th row $\bm{x}_i'$. Similarly, let $D_0 = (\by_0, \bm{\nu}_0, \bX_0, n_0)$ represent the external data set. To define the piecewise constant baseline hazard, we partition the time axis into $J$ intervals: $0 = s_0 < s_1 < \ldots < s_{J-1} < s_J = \infty$, with breakpoints selected so that each interval contains approximately the same number of events in the current data. Let $\bbeta = (\beta_1, \ldots, \beta_p)'$ denote the vector of regression coefficients, and $\blambda = (\lambda_1, \ldots, \lambda_J)'$ denote the vector of baseline hazards across the $J$ intervals.

\subsection{Prior specifications under PWEPH Model}

Under the PWEPH model, the likelihood function for the current data can be written as 
\begin{align*}
    L(\bbeta, \blambda \mid D) = \prod_{i=1}^{n} \prod_{j = 1}^{J} (\lambda_j \exp\{\bm{x}_i' \bbeta\})^{\delta_{ij} \nu_i} \exp\left\{-\delta_{ij} \left[\lambda_j (y_i - s_{j-1}) + \sum_{g=1}^{j-1}\lambda_g (s_g - s_{g-1})\right] \exp\{\bm{x}_i' \bbeta\}\right\},
\end{align*} 
where $\delta_{ij} = 1$ if subject $i$ had an event or was censored in the $j$th interval, and 0 otherwise. Table~\ref{tab:supp_PWE_summary} presents the prior specifications for each method under the PWEPH model. For both the Bayesian hierarchical model (BHM) and the commensurate prior (CP), the current and external data are modeled with PWEPH models sharing the same covariate structure but having distinct parameter vectors: $(\bbeta, \blambda)$ for the current data and $(\bbeta_0 = (\beta_{01}, \ldots, \beta_{0p})', \blambda_0 = (\lambda_{01}, \ldots, \lambda_{0J})')$ for the external data. Under the latent exchangeability prior (LEAP), the external data likelihood is modeled as a two-component mixture: one component corresponds to the current data PWEPH model and is weighted by a mixing probability $\gamma$, while the other is a distinct PWEPH model that adjusts for the same covariates but with separate parameters $(\bbeta_0, \blambda_0)$. 

\begin{table}
\centering
\caption{Summary of Prior Structures and Covariate Inclusion under the PWEPH Model}
\vspace{3mm}

\renewcommand{\arraystretch}{1.4} 
\setlength{\tabcolsep}{8pt}
\begin{tabular}{lll}
\toprule
\textbf{Method} & \textbf{Covariate Inclusion} & \textbf{Prior Specification} \\ 
\midrule

\textbf{Vague} & Outcome model & $\beta_1, \ldots, \beta_p \sim N(0, 10^2)$ and \\
& & $\lambda_1, \ldots, \lambda_J \sim N^{+}(0, 10^2)$. \\
\midrule

\textbf{PP}
& Outcome model & $\pi_{\text{PP}}(\bbeta, \blambda) \propto L(\bbeta, \blambda \mid D_0)^{a_0} ~\pi_v(\bbeta, \blambda)$, \\
& & $a_0 = 0.5$ and $\pi_v(\bbeta, \blambda)$ denotes the vague prior. \\ 
\midrule

\textbf{NPP}
& Outcome model & $\pi_{\text{NPP}}(\bbeta, \blambda, a_0) = \frac{L(\bbeta, \blambda \mid D_0)^{a_0} ~\pi_v(\bbeta, \blambda)}{Z(a_0 \mid D_0)} \cdot \pi_A(a_0)$, \\
& & $a_0 \sim \text{Uniform}(0, 1)$, \\
& & $Z(a_0 \mid D_0) = \int L(\bbeta, \blambda \mid D_0)^{a_0} ~\pi_v(\bbeta, \blambda) d(\bbeta, \blambda)$, \\
& & and $\pi_v(\bbeta, \blambda)$ denotes the vague prior. \\ 
\midrule

\textbf{PSIPP}
& Propensity score model & Applies PP within each stratum $s \in \{1, \ldots, 4\}$: 
\\ & & $\pi_{\text{PP}}(\bbeta^{s}, \blambda^{s}) \propto L(\bbeta^{s}, \blambda^{s} \mid D_{0s})^{a_{0s}} ~\pi_v(\bbeta^{s}, \blambda^{s})$. \\
\midrule

\textbf{BHM}
& Outcome model & $\lambda_k, \lambda_{0k} \sim N^{+}(0, 10^2)$ for $k = 1, \ldots, J$, \\
& & $\beta_j, \beta_{0j} \sim N(\mu_j, \tau_j^2)$ for $j = 1, \ldots, p$, \\
& & $\mu_j \sim N(0, 10^2)$ and $\tau_j \sim N^{+}(0, 0.5^2)$. \\
\midrule

\textbf{CP}
& Outcome model & $\lambda_k, \lambda_{0k} \sim N^{+}(0, 10^2)$ for $k = 1, \ldots, J$, \\
& & $\beta_{01}, \ldots, \beta_{0p} \sim N(0, 10^2)$, \\
& & $\beta_j \sim N(\beta_{0j}, \tau_j^{-1})$ for $j = 1, \ldots, p$, \\
& & $\tau_j \sim 0.1 \cdot N^{+}(200, 0.1^2) + 0.9 \cdot N^{+}(0, 5^2)$. \\
\midrule

\textbf{LEAP}
& Outcome model & $\lambda_k, \lambda_{0k} \sim N^{+}(0, 10^2)$ for $k = 1, \ldots, J$, \\
& & $\beta_j, \beta_{0j} \sim N(0, 10^2)$ for $j = 1, \ldots, p$, \\
& & $\gamma \sim \text{Uniform}(0,1)$.\\

\bottomrule
\end{tabular}
\vspace{3mm}

\caption*{The column ``Covariate Inclusion'' indicates whether baseline covariates are incorporated into the outcome model or into the propensity score model. Here, $N^{+}(\mu, \sigma^2)$ denotes a $N(\mu, \sigma^2)$ distribution truncated to the positive real line. PP = power prior; NPP = normalized power prior; PSIPP = propensity score-integrated power prior; BHM = Bayesian hierarchical model; CP = commensurate prior; LEAP = latent exchangeability prior.}

\label{tab:supp_PWE_summary}
\end{table}

\subsection{Prior specifications under CurePWEPH Model}

The CurePWEPH model assumes a common probability of cure $\pi$ for all subjects and models the survival distribution of the non-cured population using a PWEPH model. Let $f(\cdot \mid \bm{x}, \bbeta, \blambda)$ and $S(\cdot \mid \bm{x}, \bbeta, \blambda)$ denote the density and survival functions under the PWEPH model, where $\bm{x}$ denotes the covariate vector. The likelihood for the current data under the CurePWEPH model is
\begin{align*}
    L(\pi, \bbeta, \blambda \mid D) = \prod_{i=1}^{n} \Big\{ \big[(1 - \pi) f(y_i \mid \bm{x}_i, \bbeta, \blambda)\big]^{\nu_i} \cdot \big[\pi + (1 - \pi) S(y_i \mid \bm{x}_i, \bbeta, \blambda)\big]^{1 - \nu_i} \Big\}.
\end{align*} 
Table~\ref{tab:supp_CurePWE_summary} summarizes the prior specifications for all methods under the CurePWEPH model. As under the PWEPH model, both the BHM and CP specify separate CurePWEPH models for the current and external data, using the same covariate structure but distinct parameter vectors: $(\pi, \bbeta, \blambda)$ for the current data and $(\pi_{0}, \bbeta_0, \blambda_0)$ for the external data. 

Under the LEAP approach, the PWEPH component of the external data model is expressed as a two-component mixture. One component assumes exchangeability with the PWEPH structure in the current data model (sharing $\bbeta, \blambda$), while the other allows for non-exchangeability through distinct parameters $(\bbeta_0, \blambda_0)$. The components are weighted by $\gamma$ and $(1 - \gamma)$, respectively. The external data model also includes its own cure probability $\pi_{0}$. The resulting likelihood function for the external data under LEAP is
\begin{align*}
    L(\pi_0, \bbeta_0, \blambda_0 \mid D_0) = \prod_{i=1}^{n_0} \Big\{ \big[(1 - \pi_0) \tilde{f}(y_{0i})\big]^{\nu_{0i}} \cdot \big[\pi_0 + (1 - \pi_0) \tilde{S}(y_{0i})\big]^{1 - \nu_{0i}} \Big\},
\end{align*}
where $\tilde{f}(y_{0i})$ and $\tilde{S}(y_{0i})$ are given by
\begin{align*}
    \tilde{f}(y_{0i}) &= \gamma f(y_{0i} \mid \bm{x}_{0i}, \bbeta, \blambda) + (1 - \gamma) f(y_{0i} \mid \bm{x}_{0i}, \bbeta_{0}, \blambda_{0}), \\
    \tilde{S}(y_{0i}) &= \gamma S(y_{0i} \mid \bm{x}_{0i}, \bbeta, \blambda) + (1 - \gamma) S(y_{0i} \mid \bm{x}_{0i}, \bbeta_{0}, \blambda_{0}).
\end{align*}

\begin{table}
\centering
\caption{Summary of Prior Structures and Covariate Inclusion under the CurePWEPH Model}
\vspace{3mm}

\renewcommand{\arraystretch}{1.3} 
\setlength{\tabcolsep}{8pt}
\begin{tabular}{lll}
\toprule
\textbf{Method} & \textbf{Covariate Inclusion} & \textbf{Prior Specification} \\ 
\midrule

\textbf{Vague} & Outcome model & logit$(\pi) \sim N(0, 3^2),$ \\ & & $\beta_1, \ldots, \beta_p \sim N(0, 10^2)$, \\ & &
$\lambda_1, \ldots, \lambda_J \sim N^{+}(0, 10^2)$. \\
\midrule

\textbf{PP} 
 & Outcome model 
 & $\pi_{\text{PP}}(\pi,\bbeta,\blambda) \propto L(\pi,\bbeta,\blambda \mid D_0)^{a_0}\,\pi_v(\pi,\bbeta,\blambda)$, \\
 & 
 & $a_0 = 0.5$ and $\pi_v(\pi,\bbeta,\blambda)$ denotes the vague prior. \\
\midrule

\textbf{NPP} 
 & Outcome model 
 & $\pi_{\text{NPP}}(\pi,\bbeta,\blambda,a_0)
    = \dfrac{L(\pi,\bbeta,\blambda \mid D_0)^{a_0}\,\pi_v(\pi,\bbeta,\blambda)}
            {Z(a_0 \mid D_0)} \, \pi_A(a_0)$, \\
 & 
 & $a_0 \sim \text{Uniform}(0,1),$ \\
 & 
 & $Z(a_0 \mid D_0) = \int L(\pi,\bbeta,\blambda \mid D_0)^{a_0}\,
                     \pi_v(\pi,\bbeta,\blambda)\, d(\pi,\bbeta,\blambda),$ \\
 & 
 & and $\pi_v(\pi,\bbeta,\blambda)$ denotes the vague prior. \\
\midrule

\textbf{PSIPP} 
& Propensity score model & Applies PP within each stratum $s \in \{1, \ldots, 4\}$: 
\\ & & $\pi_{\text{PP}}(\pi^{s}, \bbeta^{s}, \blambda^{s}) \propto L(\pi^{s}, \bbeta^{s}, \blambda^{s} \mid D_{0s})^{a_{0s}} ~\pi_v(\pi^{s}, \bbeta^{s}, \blambda^{s})$. \\
\midrule

\textbf{BHM}
& Outcome model & $\text{logit}(\pi), \text{logit}(\pi_{0}) \sim N(0, 3^2),$ \\ & & $\lambda_k, \lambda_{0k} \sim N^{+}(0, 10^2)$ for $k = 1, \ldots, J$, \\
& & $\beta_j, \beta_{0j} \sim N(\mu_j, \tau_j^2)$ for $j = 1, \ldots, p$, \\
& & $\mu_j \sim N(0, 10^2)$ and $\tau_j \sim N^{+}(0, 0.5^2)$. \\
\midrule

\textbf{CP} 
& Outcome model & $\text{logit}(\pi), \text{logit}(\pi_{0}) \sim N(0, 3^2),$ \\ & & $\lambda_k, \lambda_{0k} \sim N^{+}(0, 10^2)$ for $k = 1, \ldots, J$, \\
& & $\beta_{01}, \ldots, \beta_{0p} \sim N(0, 10^2)$, \\
& & $\beta_j \sim N(\beta_{0j}, \tau_j^{-1})$ for $j = 1, \ldots, p$, \\
& & $\tau_j \sim 0.1 \cdot N^{+}(200, 0.1^2) + 0.9 \cdot N^{+}(0, 5^2)$. \\
\midrule

\textbf{LEAP}
& Outcome model & $\text{logit}(\pi), \text{logit}(\pi_{0}) \sim N(0, 3^2),$ \\ & & $\lambda_k, \lambda_{0k} \sim N^{+}(0, 10^2)$ for $k = 1, \ldots, J$, \\
& & $\beta_j, \beta_{0j} \sim N(0, 10^2)$ for $j = 1, \ldots, p$, \\
& & $\gamma \sim \text{Uniform}(0,1)$.\\

\bottomrule
\end{tabular}
\vspace{3mm}

\caption*{The column ``Covariate Inclusion'' indicates whether baseline covariates are incorporated into the outcome model or into the propensity score model. Here, $N^{+}(\mu, \sigma^2)$ denotes a $N(\mu, \sigma^2)$ distribution truncated to the positive real line. PP = power prior; NPP = normalized power prior; PSIPP = propensity score-integrated power prior; BHM = Bayesian hierarchical model; CP = commensurate prior; LEAP = latent exchangeability prior.}

\label{tab:supp_CurePWE_summary}
\end{table}
\FloatBarrier

\subsection{Estimation of marginal two-year RFS probabilities}

We now describe the Bayesian bootstrap procedure used to compute marginal two-year RFS probabilities and the corresponding treatment effect $\Delta$ in the survival analysis. For each survival model and prior specification (excluding the PSIPP), the procedure proceeds as follows:

\begin{enumerate}
    \item Split the current and external data sets by treatment arm $a \in \{0 = \text{Control}, 1 = \text{IFN}\}$.

    \item For each treatment arm $a$ in the current data, fit the specified survival model (PWEPH or CurePWEPH) under the chosen prior and obtain posterior samples of the model parameters $\btheta_a$. When applicable, the prior incorporates information from the corresponding treatment arm in the external data.

    \item For each posterior draw of $\btheta_a$:
    \begin{enumerate}
        \item Compute the conditional two-year RFS probability for each subject $i = 1, \ldots, n$ in the current data under treatment $a$, denoted by $\hat{S}_a(2 \mid \bm{x}_i, \btheta_a)$, using the subject's baseline covariates $\bm{x}_i$.

        \item Draw a vector of weights $\bm{w} = (w_1, \ldots, w_{n})' \sim \text{Dirichlet}(1, \ldots, 1)$ and form the marginal two-year RFS probability as $$\hat{S}_a(2) = \sum_{i=1}^{n} w_i \cdot \hat{S}_a(2 \mid \bm{x}_i, \btheta_a).$$

        \item Compute the treatment effect as $\Delta = \hat{S}_1(2) - \hat{S}_0(2)$.
    \end{enumerate}

    \item Repeat step (3) across posterior draws to obtain the posterior distribution of $\Delta$.  
\end{enumerate}

In this procedure, baseline covariates from all subjects (regardless of treatment assignment) are used to estimate two-year RFS probabilities for both treatment arms, thereby marginalizing over a common covariate distribution. Because the covariates were measured at baseline (i.e., pre-treatment), the underlying covariate distribution is naturally shared across treatment groups. 

For the PSIPP, we apply a similar procedure with modifications to account for stratification. Specifically, in step (2), a simplified survival model without baseline covariates is fit separately within each propensity score stratum $k \in \{1, \ldots, 4\}$, yielding posterior samples of stratum-specific parameters $\btheta_{a, k}$. For each posterior draw, the two-year RFS probability within stratum $k$ is computed as $\hat{S}_{a, k}(2 \mid \btheta_{a,k})$. The marginal two-year RFS probability for treatment group $a$ is then obtained by averaging across strata, 
$$\hat{S}_a(2) = \frac{1}{4} \sum_{k = 1}^4 \hat{S}_{a, k}(2 \mid \btheta_{a,k}).$$
Because the strata are defined by the quartiles of the PS among the current-study subjects, each stratum contains approximately one quarter of these subjects, and the equal weights average over the covariate distribution of the current study.
Repeating this calculation across posterior draws yields the posterior distribution of $\Delta$ under the PSIPP.

\section{Additional results from survival analysis}

\subsection{Selection of the number of intervals via ELPD}\label{appendix:elpd_plot}
\begin{figure}
    \centering
    \includegraphics[width=0.95\linewidth]{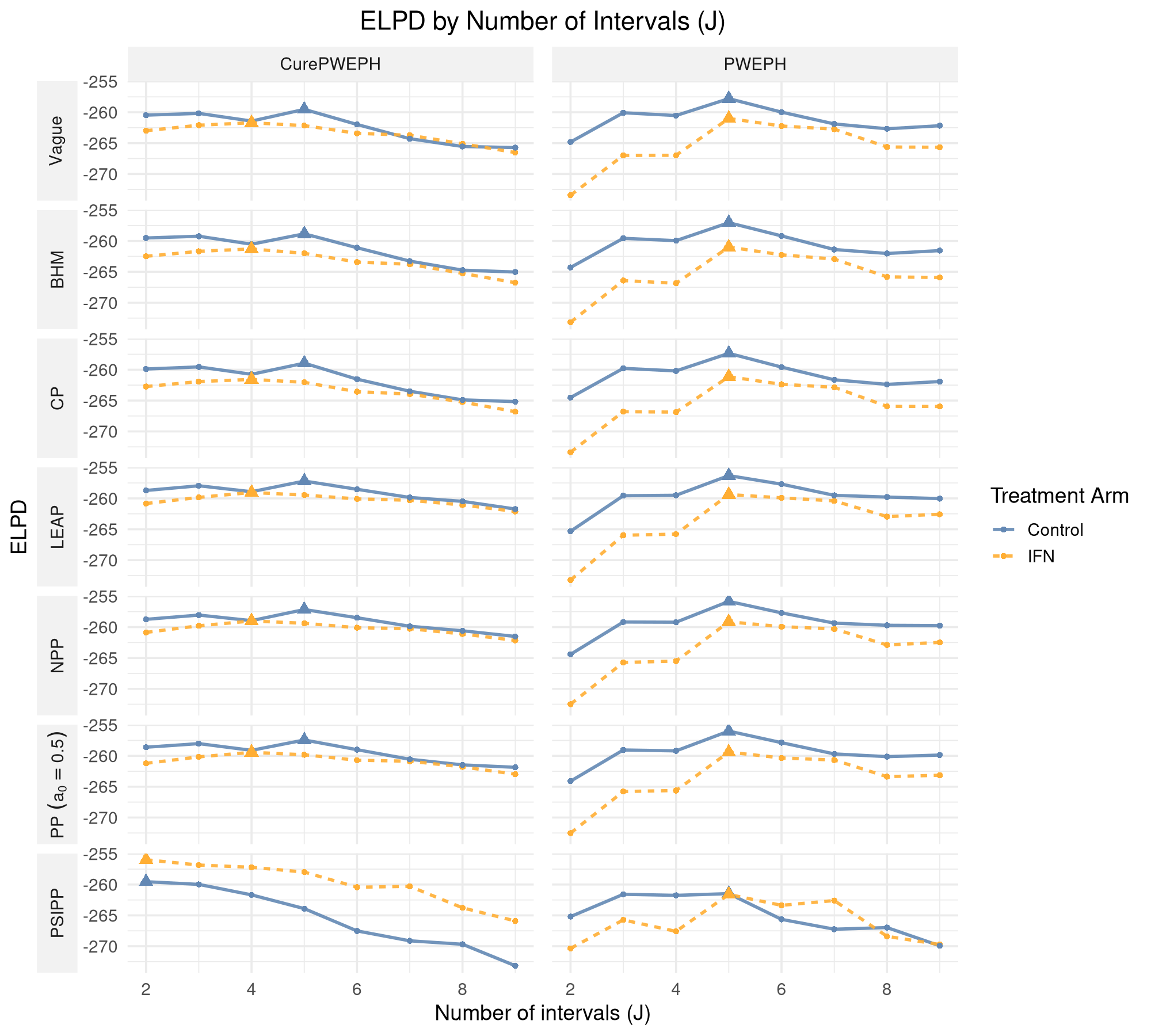}
    \caption{\textbf{Expected log predictive density (elpd) as a function of the number of intervals (J), stratified by prior (rows) and model type (columns; CurePWEPH vs. PWEPH)}. Within each panel, treatment arms are distinguished by color and line type. Triangles indicate the number of intervals at which each prior–model combination attains its highest elpd value. BHM = Bayesian hierarchical model; CP = commensurate prior; LEAP = latent exchangeability prior; NPP = normalized power prior; PP = power prior; PSIPP = propensity score-integrated power prior.}
    \label{fig:elpd_plots}
\end{figure}

Figure \ref{fig:elpd_plots} displays the expected log predictive density (elpd) values across different numbers of intervals ($J = 2$ to 9), stratified by prior and model type (CurePWEPH or PWEPH) for both treatment arms. These panels illustrate how predictive performance varies with increasing model flexibility and allow us to identify the optimal number of intervals for each combination of prior and model type. For the PWEPH model, the highest elpd values were consistently attained at $J = 5$ across both treatment arms and all priors considered. In contrast, the CurePWEPH model showed greater variability. Under the PSIPP, the highest elpd value was achieved at $J = 2$ for both treatment arms, indicating a preference for a simpler specification. Under the other priors (BHM, CP, LEAP, NPP and PP), the optimal number of intervals was $J = 4$ for the IFN arm and $5$ for the control arm. 

\subsection{Effective number of patients borrowed from the external study}\label{appendix:ess_plot_heatmap}
\begin{figure}
    \centering
    \includegraphics[width=\linewidth]{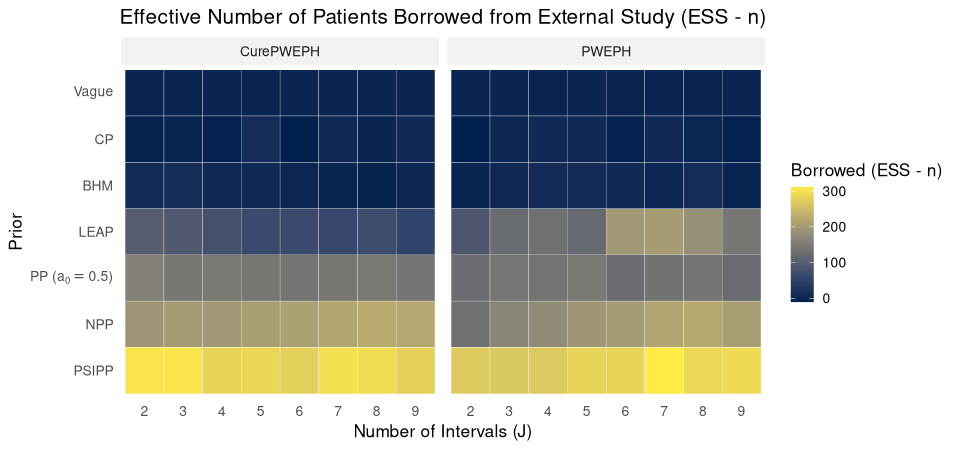}
    \caption{\textbf{Effective number of patients borrowed from external data ($\text{ESS} - n$) by number of intervals (J), stratified by prior (rows) and model type (columns; CurePWEPH vs. PWEPH)}. Within each panel, the amount of borrowing from the external data is represented by color, with darker blue indicating less borrowing and yellow indicating more borrowing. BHM = Bayesian hierarchical model; CP = commensurate prior; LEAP = latent exchangeability prior; NPP = normalized power prior; PP = power prior; PSIPP = propensity score-integrated power prior.}
    \label{fig:ess_plot_heatmap}
\end{figure}

To assess the degree of information borrowing from the external study under different prior specifications, we examined the effective number of patients borrowed, defined as $(\text{ESS} - n)$, where ESS denotes the effective sample size, and $n$ is the number of individuals in the current trial. Figure \ref{fig:ess_plot_heatmap} displays the amount of information borrowing across values of the number of intervals ($J$) for each combination of prior and model type (CurePWEPH or PWEPH), with darker blue indicating less borrowing and yellow indicating more borrowing from the external study. Among the informative priors, PSIPP consistently incorporates the largest amount of external information across both model types and all values of $J$, followed by NPP. In contrast, CP and BHM show only minimal borrowing. The degree of borrowing is also relatively stable across different choices of $J$.

\FloatBarrier

\section{Additional results from ATE estimation}

\begin{table}[htbp]
\caption{\textbf{Average treatment effect estimates.} The first two rows display the estimates of the ATE using maximum likelihood: the first with intercept and treatment, and the second with all the covariates (including the intercept).
We also report the standard error and the $95\%$ confidence interval.
Lastly, we present the posterior mean, standard deviation and $95\%$ credibility interval for the ATE estimated by BMA.
We use uncertainty interval (UI) to stand for either a CI or a BCI.}
\centering
\begin{tabular}{lccc}
  \hline
    Model & Estimate & SE/SD & $95\%$ UI \\ 
  \hline
  GLM - treatment only & $-0.0296$ & $0.0349$ & $(-0.098, 0.039)$ \\
  GLM - full & $-0.0045$ & $0.0337$ & $(-0.071, 0.061)$ \\ 
  NPP (BMA) & $-0.0389$ & $0.0208$ & $(-0.080, 0.003)$ \\
  \hline
\end{tabular}
\label{tab:ate_glm}
\end{table}

\begin{figure}[htbp]
    \centering
    \includegraphics[scale=.5]{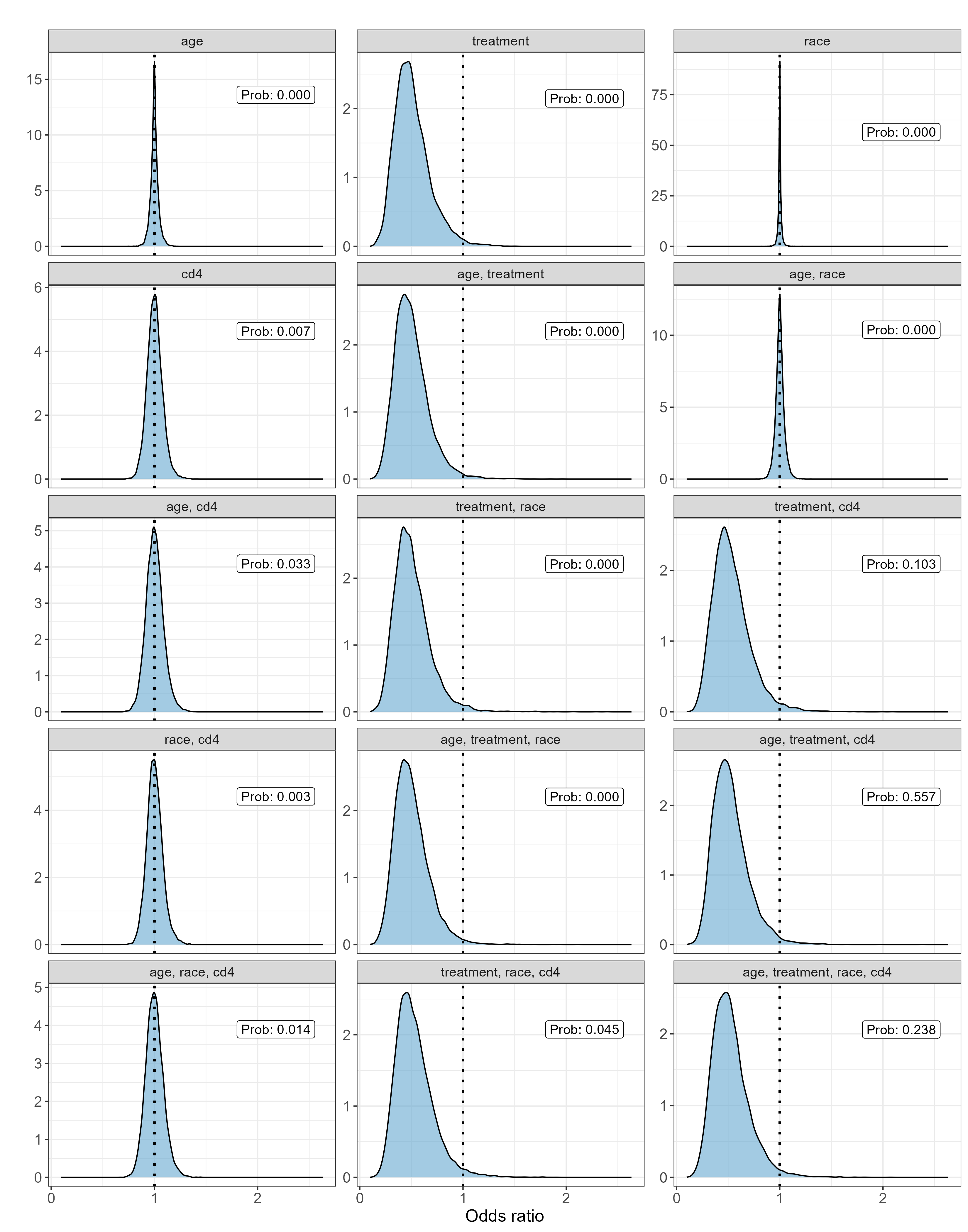}
    \caption{\textbf{Marginalized odds ratio across models}.
    We show the posterior density for the odds ratio $\frac{\mu_1/(1 - \mu_1)}{\mu_0/(1-\mu_0)}$ between treatment and control for each combination of covariates (i.e. each model) along with the estimated posterior probability for each model.
    Dotted vertical line marks OR = 1.}
    \label{fig:or_posterior_bymodel}
\end{figure}

\begin{figure}[htbp]
    \centering
    \includegraphics[scale=.5]{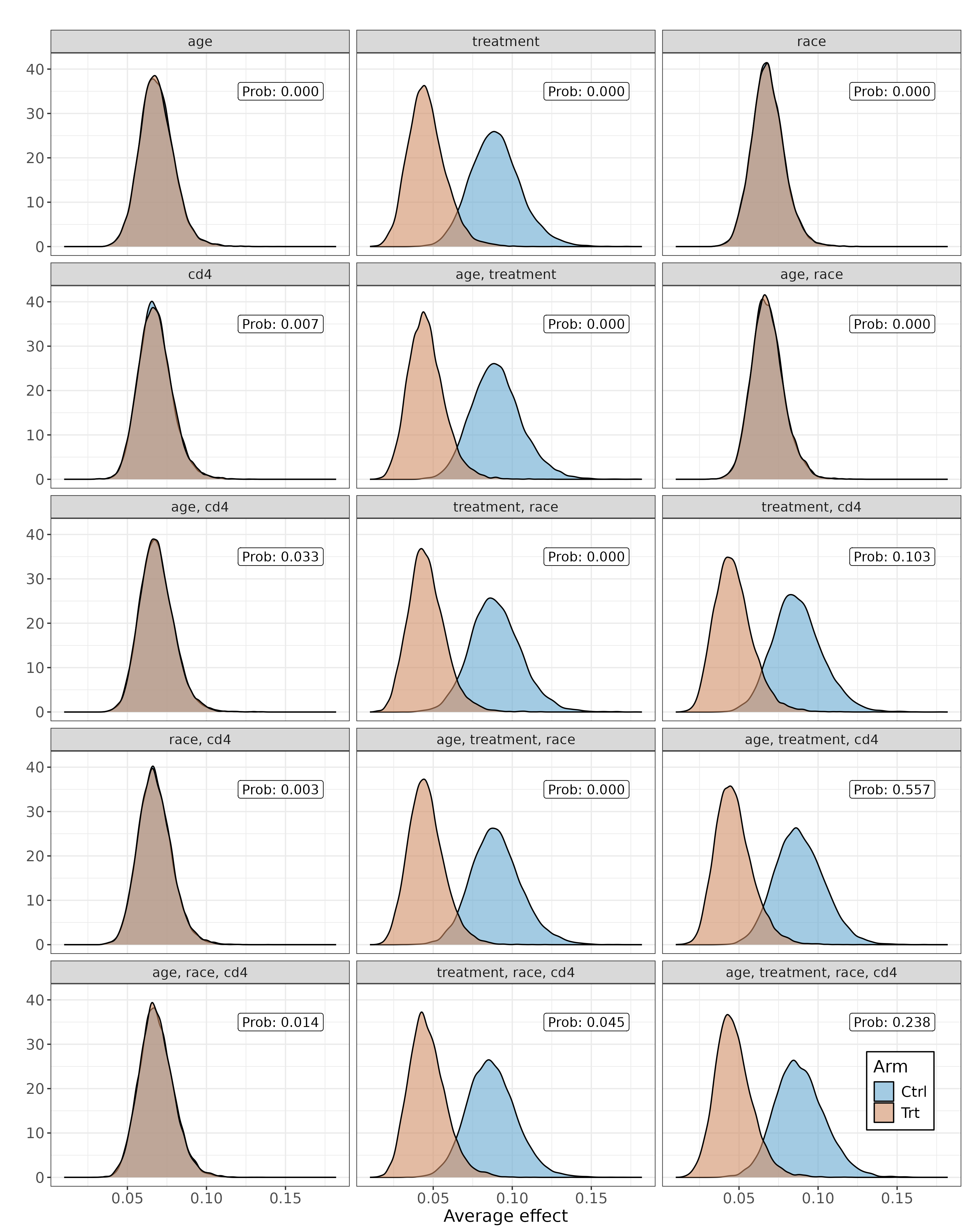}
    \caption{\textbf{Posterior distributions of the marginal means per arm over all models}.    We show the posterior distribution of the marginal expected values $\mu_1$ and $\mu_0$ over each of the models considered, and the estimates are obtained \textit{via} Bayesian bootstrap (see text for details). 
    We also show the estimated posterior probability for each model.}
    \label{fig:average_effect_posterior_bymodel}
\end{figure}

\section{Supplementary figures}

\begin{figure}[htbp]
    \centering
    \includegraphics[scale=.45]{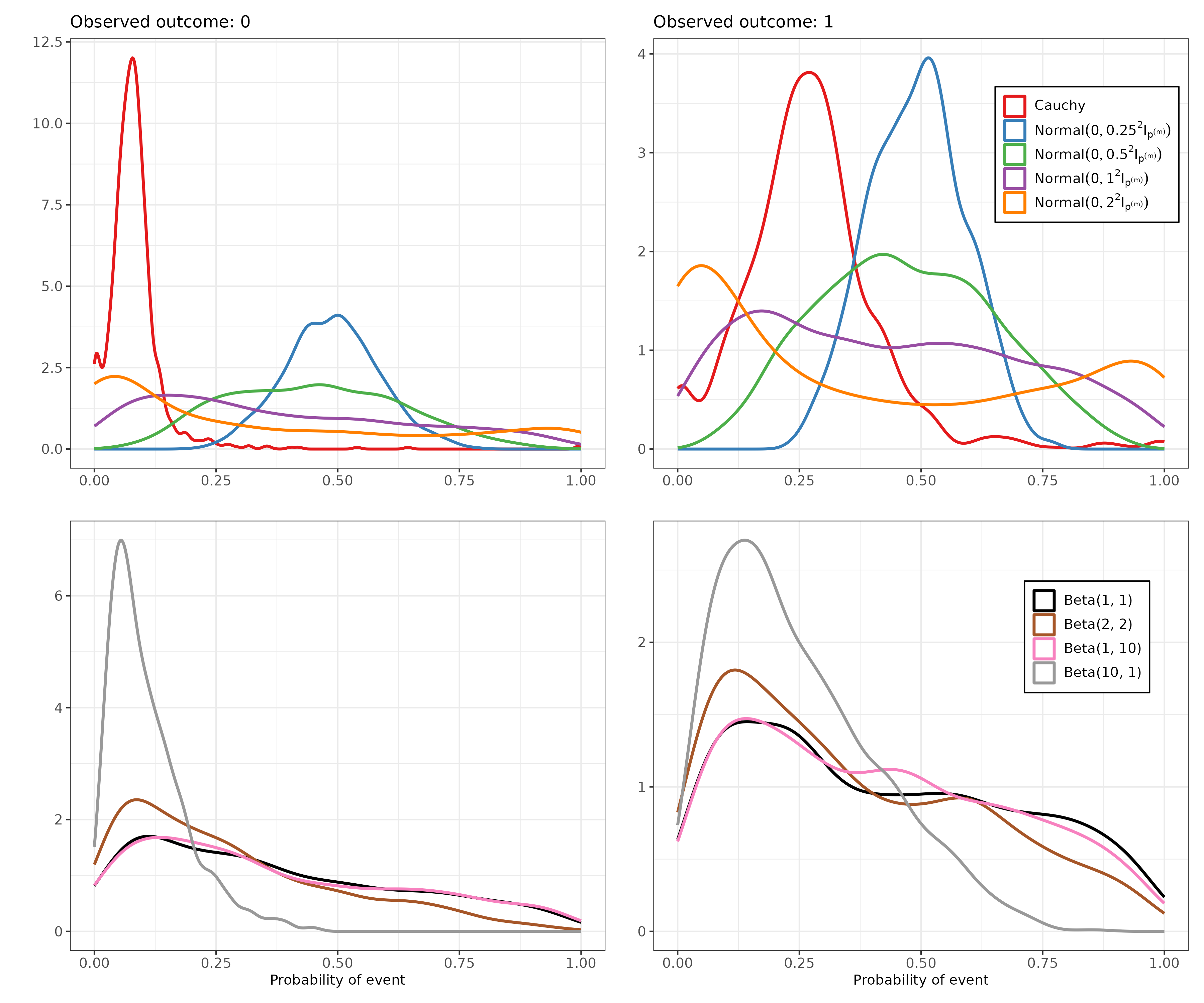}
    \caption{\textbf{Prior predictive distribution for the outcome under different priors.}
    We show the prior predictive probability for two subjects (one with outcome=0 and another with outcome=1) for the different initial priors considered here under the normalized power prior (NPP) model.
    In the top panels, line colors show the choice of initial prior and in the bottom panel, the prior on the discounting parameter.
    The weakly informative prior (WIP) was specified to be Cauchy$(0, 10)$ for the intercept and Cauchy$(0, 2.5)$ for the other coefficients.
    The results for the top panels were obtained using a Beta$(1, 1)$ prior on the discounting parameter (black in the bottom panels).
    For the bottom panels, results were obtained by using the Normal$(0, \boldsymbol{I})$ initial prior on the coefficients (purple in the top panels).
    }
    \label{fig:ppc_all}
\end{figure}

\begin{figure}[htbp]
    \centering
    \includegraphics[scale=.55]{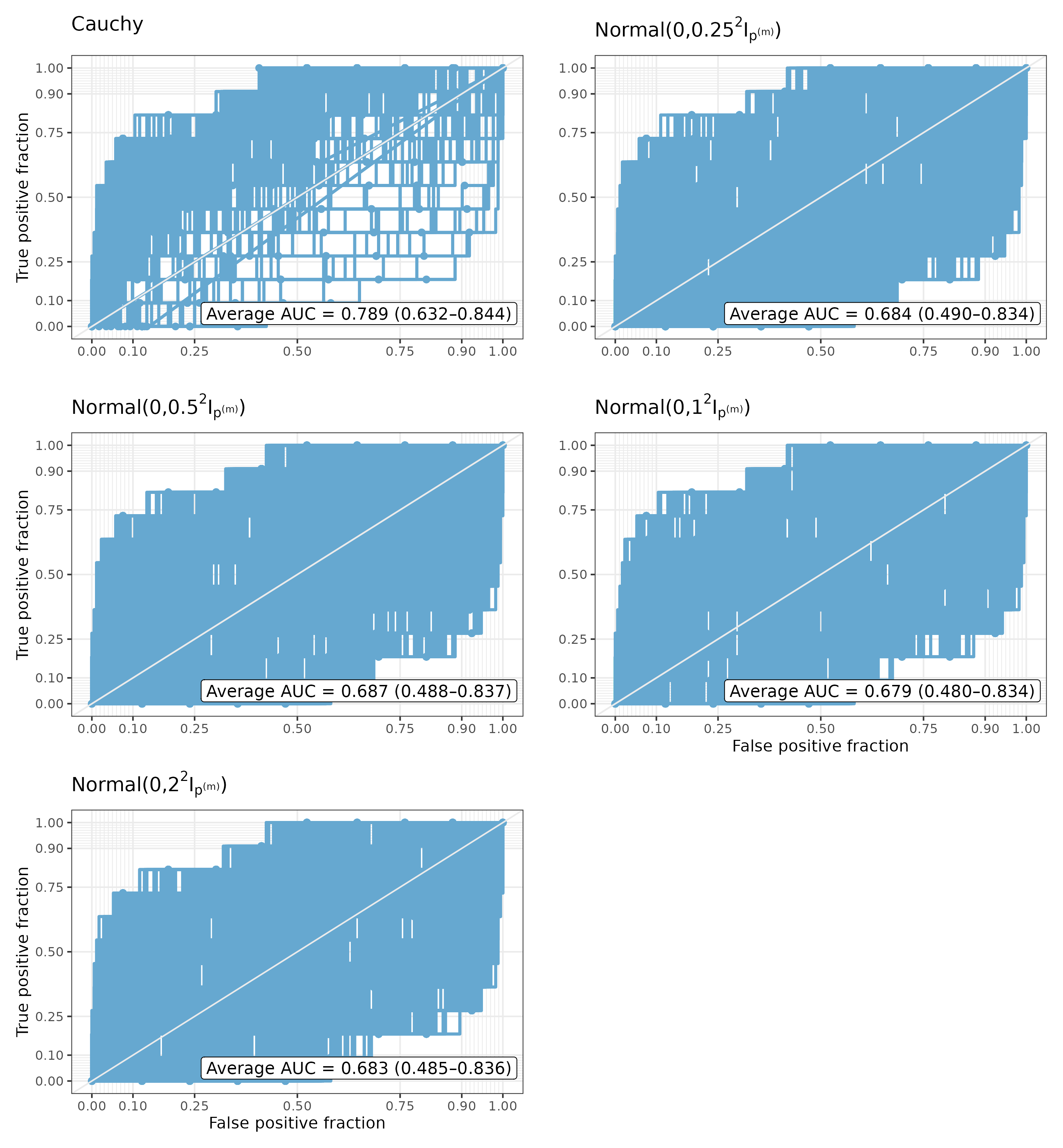}
    \caption{\textbf{Prior predictive ROC (pROC) curves under different priors on $\bbeta$.} Using the \textbf{current data} as the gold standard, we show the ROC curves implied by each initial prior and the historical data.
    Each light blue curve is a realization from the prior predictive and we mark the $x=y$ line as the naive classification model.
    We also show the posterior average AUC and its $90\%$ credibility interval.}
    \label{fig:roc_beta}
\end{figure}

\begin{figure}[htbp]
    \centering
    \includegraphics[scale=.55]{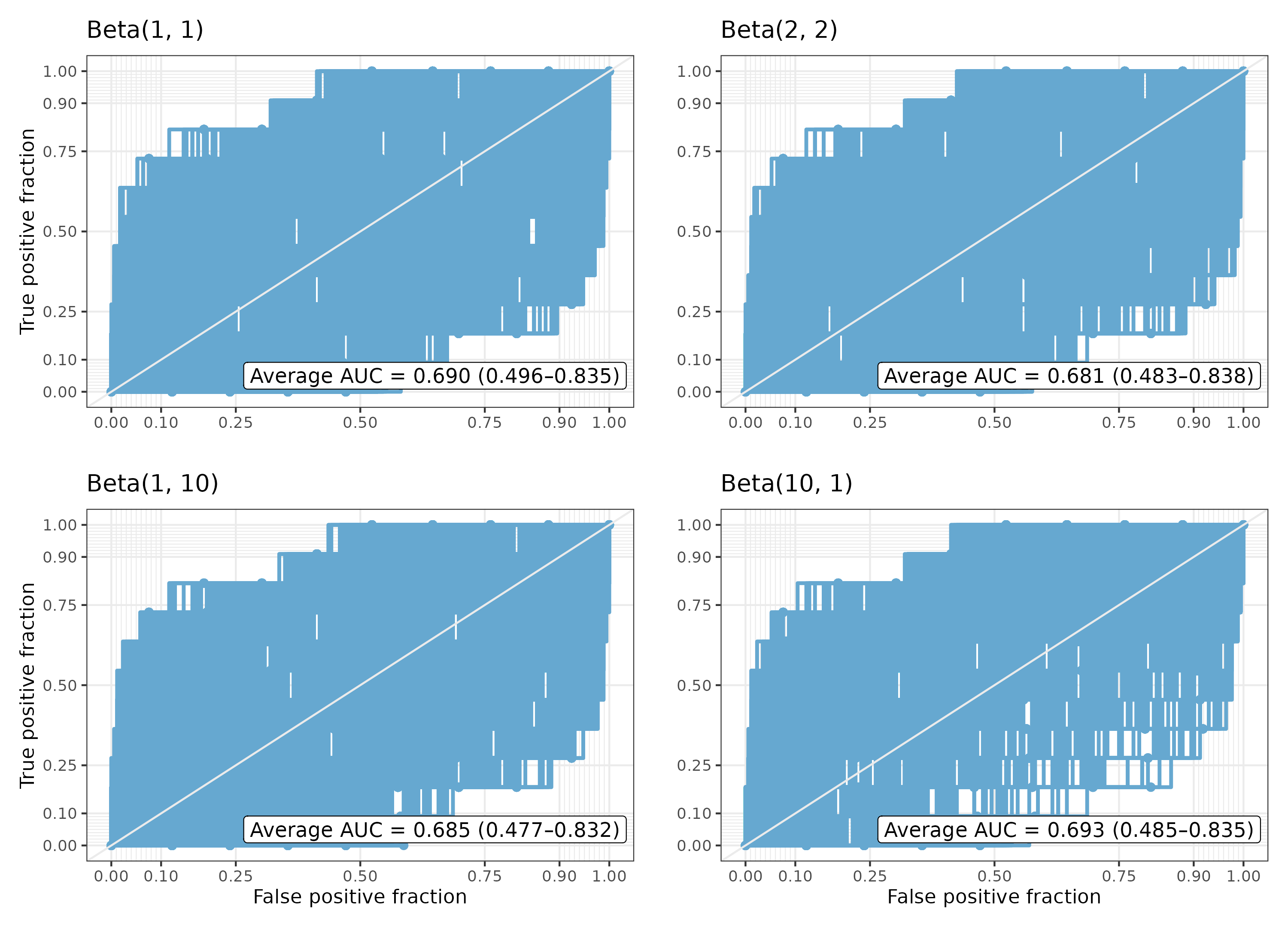}
    \caption{\textbf{Prior predictive ROC curves under different priors on $a_0$.} Under a containment (normal with mean zero and standard deviation 1) initial prior on the coefficients. In the box we present the average AUC and its $90\%$ credibility interval.}
    \label{fig:roc_a0}
\end{figure}

\begin{figure}[htbp]
    \centering
    \includegraphics[width=0.95\linewidth]{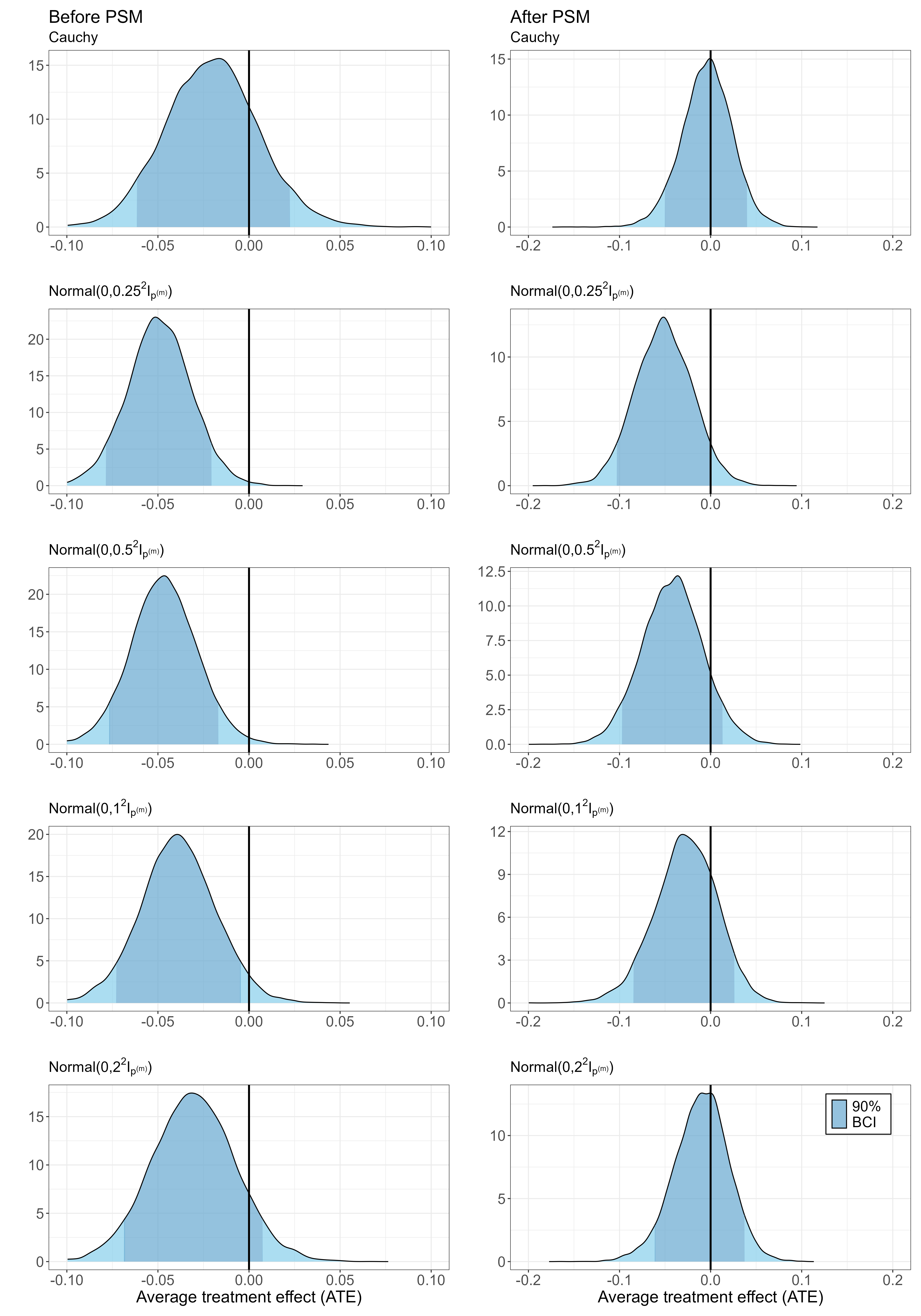}
    \caption{\textbf{Comparison of ATE distributions before and after PSM.} We show the ATE distribution for each choice of initial prior as in Figure \ref{fig:ate_priors} where each column presents the results before and after PSM.}
    \label{fig:ates_before_after_PSM}
\end{figure}

\begin{figure}[htbp]
     \centering
     \includegraphics[scale=.4]{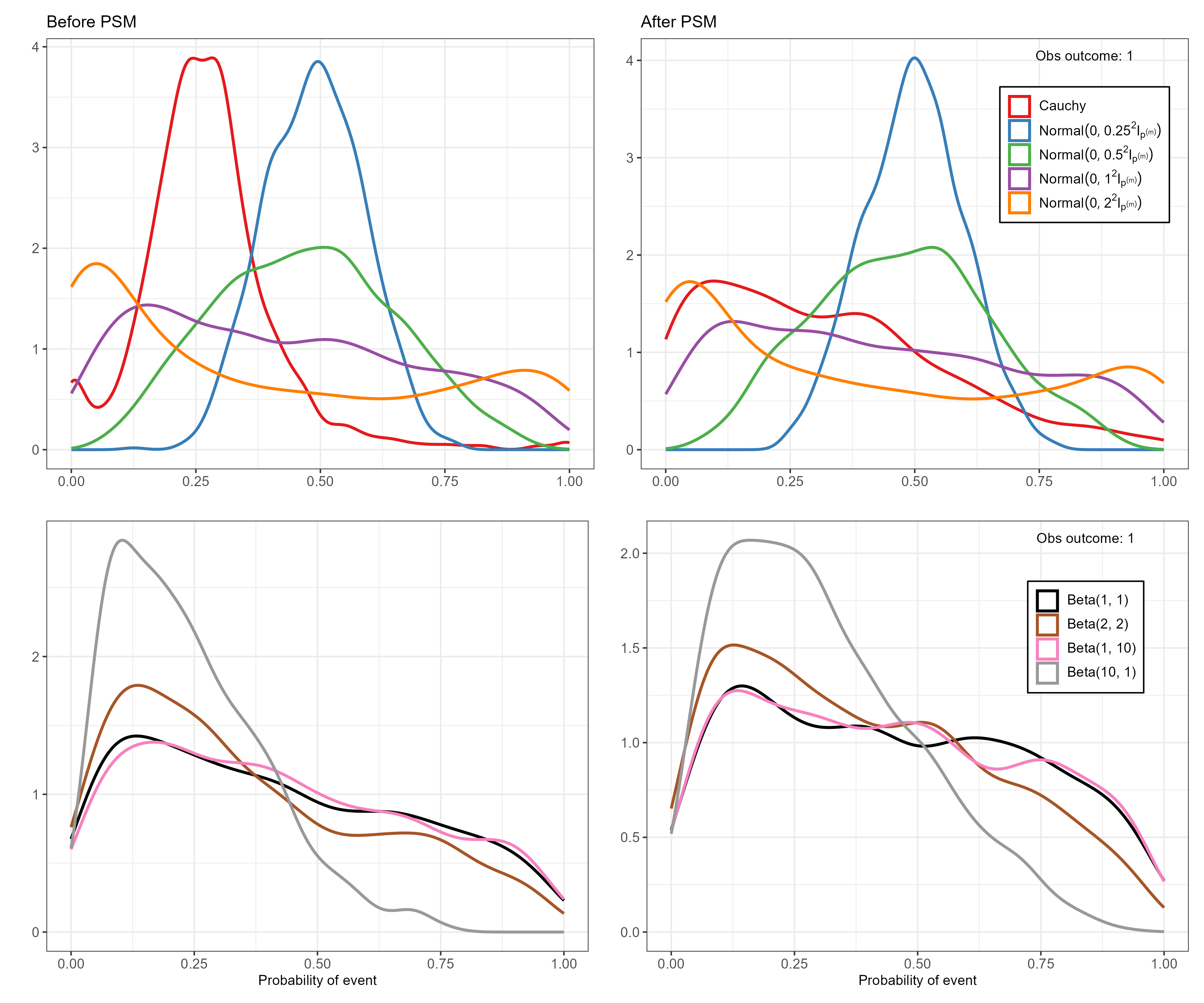}
    \caption{\textbf{Comparison of prior predictive distributions under different priors for one subject before and after PSM.} Here we show the prior predictive probability as in Figure \ref{fig:ppc_all} for one subject with outcome $= 1$.}
    \label{fig:ppc_before_after_PSM}
\end{figure}

\begin{figure}[htbp]
    \centering
    \includegraphics[width=0.95\linewidth]{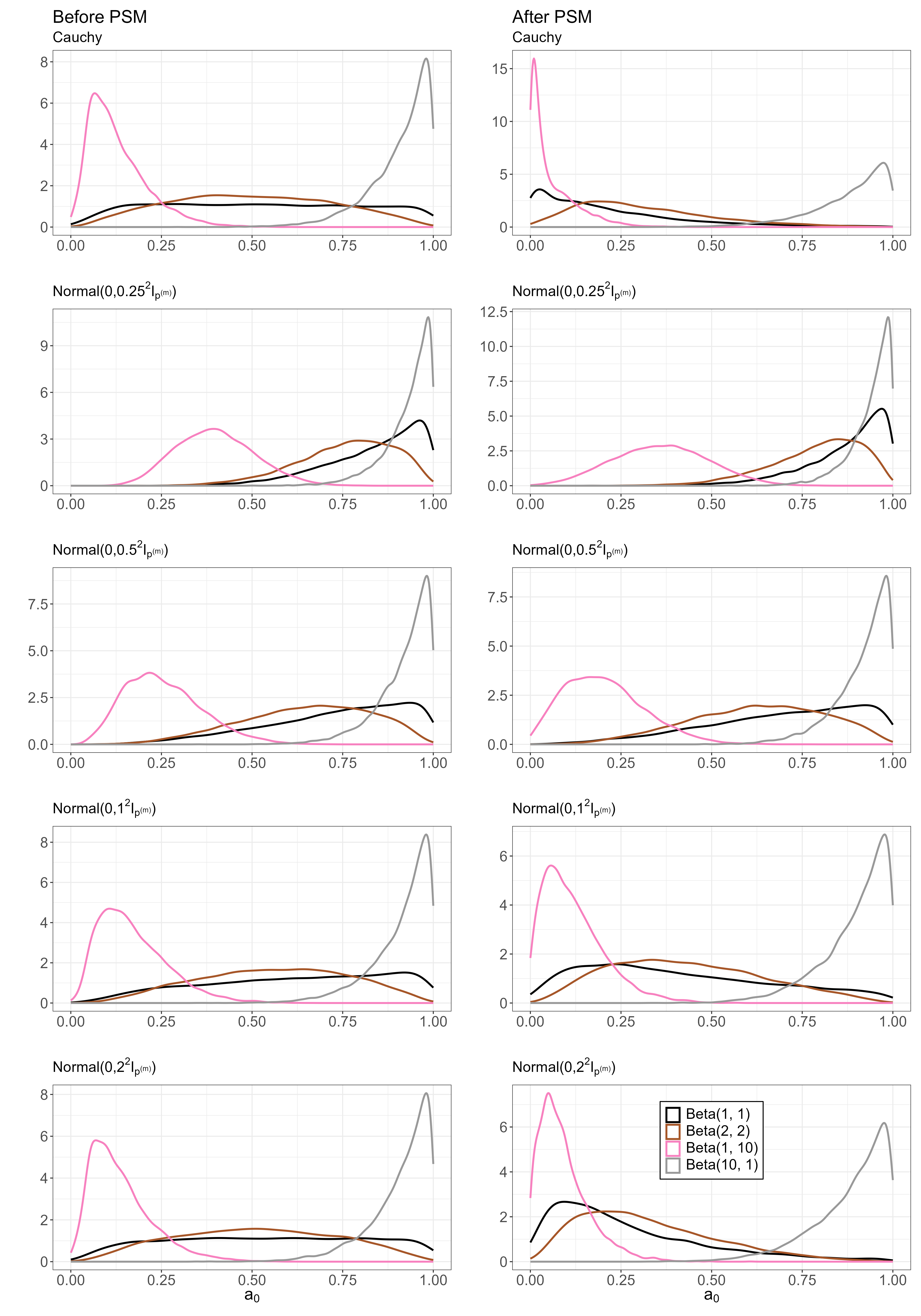}
    \caption{\textbf{Posterior distributions for $a_0$ under different priors.} We show the posterior distribution for different priors on $a_0$ and $\bbeta$. Each column corresponds to the results before and after PSM. Cauchy prior was specified as previously described.}
    \label{fig:post_a0}
\end{figure}